\documentclass[preprint,showpacs,preprintnumbers,amsmath,amssymb]{revtex4}

\usepackage[usenames]{color}

\usepackage{graphicx}% Include figure files
\usepackage{dcolumn}% Align table columns on decimal point
\usepackage{bm}% bold math

\def\lsim{\mathrel{\rlap{\lower4pt\hbox{\hskip1pt$\sim$}}
    \raise1pt\hbox{$<$}}}                % less than or approx. symbol
\def\gsim{\mathrel{\rlap{\lower4pt\hbox{\hskip1pt$\sim$}}
    \raise1pt\hbox{$>$}}}                % greater than or approx. symbol

\begin{document}

\preprint{CDFNOTE 7475}

\title{A Search for Supersymmetric Higgs Bosons in
the Di-Tau Decay Mode in $p\bar{p}$ Collisions at $\sqrt{s}=1.8$ TeV}% Force line breaks with \\
%\author{The CDF Collaboration}
% \altaffiliation[Also at ]{Physics Department, XYZ University.}%Lines break automatically or can be forced with \\
%\author{Second Author}%
% \email{Second.Author@institution.edu}
%\affiliation{%
%Authors' institution and/or address\\
%This line break forced with \textbackslash\textbackslash
%}%

%\date{\today}% It is always \today, today,
             %  but any date may be explicitly specified
\date{June 13, 2005}

\begin{abstract}
A search for direct production of Higgs bosons in the di-tau decay mode is performed 
with $86.3 \pm 3.5$ pb$^{-1}$ of data collected with the Collider Detector at Fermilab 
during the 1994$-$1995 data taking period of the Tevatron. We search for events where one 
tau decays to an electron plus neutrinos and the other tau decays hadronically.  
We perform a counting experiment and set limits on the cross section 
for supersymmetric 
Higgs boson production 
where $\tan{\beta}$ is large and $m_{A}$ is small.  
For a benchmark parameter space point where $m_{A^0}=100$ GeV/$c^2$ and $\tan{\beta}=50$, 
we limit the production cross section multiplied by the branching ratio
to be less than 77.9 pb 
 at the 95\% confidence level 
compared to the theoretically predicted value of 11.0 pb.  
This is the first search for Higgs bosons decaying to tau pairs at a hadron collider.

%Although the limits reported for this analysis are not as 
%stringent as those previously set by a CDF search for
%SUSY Higgs events with a 4-b final state, this analysis does 
%utilize a di-tau mass reconstruction technique for the
% first time using hadron collider data.  
%The mass reconstruction technique requires that the taus not be
%back-to-back in the azimuthal plane.  In this analysis, 
%the counting experiment using both back-to-back and non-back-to-back
%events
%gives stronger limits than the one using the 
%mass reconstruction technique
%due mainly to the utilization of a larger sample size in the 
%counting experiment; the mass
%resolution is not adequate to compensate for the reduced statistics
%when the back-to-back requirement is made.
\end{abstract}

% PACS, the Physics and Astronomy 

\maketitle

\font\eightit=cmti8
\def\r#1{\ignorespaces $^{#1}$}
\hfilneg
\begin{sloppypar}
\noindent
D.~Acosta,\r {14} T.~Affolder,\r 7 M.G.~Albrow,\r {13} D.~Ambrose,\r {36}   
D.~Amidei,\r {27} K.~Anikeev,\r {26} J.~Antos,\r 1 
G.~Apollinari,\r {13} T.~Arisawa,\r {50} A.~Artikov,\r {11}
% ...
\end{sloppypar}
\vskip .026in
\begin{center}
(CDF Collaboration)
\end{center}

\vskip .026in
\begin{center}
\r 1  {\eightit Institute of Physics, Academia Sinica, Taipei, Taiwan 11529, 
Republic of China} \\
\r 2  {\eightit Argonne National Laboratory, Argonne, Illinois 60439} \\
\r 3  {\eightit Istituto Nazionale di Fisica Nucleare, University of Bologna,
I-40127 Bologna, Italy} \\
\r 4  {\eightit Brandeis University, Waltham, Massachusetts 02254} \\
\r 5  {\eightit University of California at Davis, Davis, California  95616} \\
\r 6  {\eightit University of California at Los Angeles, Los 
Angeles, California  90024} \\ 
\r 7  {\eightit University of California at Santa Barbara, Santa Barbara, California 
93106} \\ 
\r 8 {\eightit Instituto de Fisica de Cantabria, CSIC-University of Cantabria, 
39005 Santander, Spain} \\
\r 9  {\eightit Carnegie Mellon University, Pittsburgh, Pennsylvania  15213} \\
\r {10} {\eightit Enrico Fermi Institute, University of Chicago, Chicago, 
Illinois 60637} \\
\r {11}  {\eightit Joint Institute for Nuclear Research, RU-141980 Dubna, Russia}
\\
\r {12} {\eightit Duke University, Durham, North Carolina  27708} \\
\r {13} {\eightit Fermi National Accelerator Laboratory, Batavia, Illinois 
60510} \\
\r {14} {\eightit University of Florida, Gainesville, Florida  32611} \\
\r {15} {\eightit Laboratori Nazionali di Frascati, Istituto Nazionale di Fisica
               Nucleare, I-00044 Frascati, Italy} \\
\r {16} {\eightit University of Geneva, CH-1211 Geneva 4, Switzerland} \\
\r {17} {\eightit Glasgow University, Glasgow G12 8QQ, United Kingdom}\\
\r {18} {\eightit Harvard University, Cambridge, Massachusetts 02138} \\
\r {19} {\eightit Hiroshima University, Higashi-Hiroshima 724, Japan} \\
\r {20} {\eightit University of Illinois, Urbana, Illinois 61801} \\
\r {21} {\eightit The Johns Hopkins University, Baltimore, Maryland 21218} \\
\r {22} {\eightit Institut f\"{u}r Experimentelle Kernphysik, 
Universit\"{a}t Karlsruhe, 76128 Karlsruhe, Germany} \\
\r {23} {\eightit Center for High Energy Physics: Kyungpook National
University, Taegu 702-701; Seoul National University, Seoul 151-742; and
SungKyunKwan University, Suwon 440-746; Korea} \\
\r {24} {\eightit Ernest Orlando Lawrence Berkeley National Laboratory, 
Berkeley, California 94720} \\
\r {25} {\eightit University College London, London WC1E 6BT, United Kingdom} \\
\r {26} {\eightit Massachusetts Institute of Technology, Cambridge,
Massachusetts  02139} \\   
\r {27} {\eightit University of Michigan, Ann Arbor, Michigan 48109} \\
\r {28} {\eightit Michigan State University, East Lansing, Michigan  48824} \\
\r {29} {\eightit Institution for Theoretical and Experimental Physics, ITEP,
Moscow 117259, Russia} \\
\r {30} {\eightit University of New Mexico, Albuquerque, New Mexico 87131} \\
\r {31} {\eightit Northwestern University, Evanston, Illinois  60208} \\
\r {32} {\eightit The Ohio State University, Columbus, Ohio  43210} \\
\r {33} {\eightit Osaka City University, Osaka 588, Japan} \\
\r {34} {\eightit University of Oxford, Oxford OX1 3RH, United Kingdom} \\
\r {35} {\eightit Universita di Padova, Istituto Nazionale di Fisica 
          Nucleare, Sezione di Padova, I-35131 Padova, Italy} \\
\r {36} {\eightit University of Pennsylvania, Philadelphia, 
        Pennsylvania 19104} \\   
\r {37} {\eightit Istituto Nazionale di Fisica Nucleare, University and Scuola
               Normale Superiore of Pisa, I-56100 Pisa, Italy} \\
\r {38} {\eightit University of Pittsburgh, Pittsburgh, Pennsylvania 15260} \\
\r {39} {\eightit Purdue University, West Lafayette, Indiana 47907} \\
\r {40} {\eightit University of Rochester, Rochester, New York 14627} \\
\r {41} {\eightit Rockefeller University, New York, New York 10021} \\
\r {42} {\eightit Instituto Nazionale de Fisica Nucleare, Sezione di Roma,
University di Roma I, ``La Sapienza," I-00185 Roma, Italy}\\
\r {43} {\eightit Rutgers University, Piscataway, New Jersey 08855} \\
\r {44} {\eightit Texas A\&M University, College Station, Texas 77843} \\
\r {45} {\eightit Texas Tech University, Lubbock, Texas 79409} \\
\r {46} {\eightit Institute of Particle Physics, University of Toronto, Toronto
M5S 1A7, Canada} \\
\r {47} {\eightit Istituto Nazionale di Fisica Nucleare, University of Trieste/\
Udine, Italy} \\
\r {48} {\eightit University of Tsukuba, Tsukuba, Ibaraki 305, Japan} \\
\r {49} {\eightit Tufts University, Medford, Massachusetts 02155} \\
\r {50} {\eightit Waseda University, Tokyo 169, Japan} \\
\r {51} {\eightit University of Wisconsin, Madison, Wisconsin 53706} \\
\r {52} {\eightit Yale University, New Haven, Connecticut 06520} \\
\end{center}

                             % Classification Scheme.
%\keywords{Suggested keywords}%Use showkeys class option if keyword

%\author{Charlie Author}
% \homepage{http://www.Second.institution.edu/~Charlie.Author}
%\affiliation{
%Second institution and/or address\\
%This line break forced% with \\
%}%

%\include{overview} 
%\include{selection} 
%\include{backgrounds}
%\include{systematics}
%\include{results}
%\include{conclusions}
%\include{acknowledgements}

\section{\label{sec:Introduction}Introduction}

The Higgs mechanism in the Minimal Supersymmetric Standard Model (MSSM)~\cite{mssm1,mssm2,mssm3} provides a way to assign a mass to each particle 
while preserving the gauge invariance of the theory just as in the Standard Model (SM).  
%The SM introduces a single SU(2) Higgs doublet and stipulates that one physical scalar particle should exist (denoted $h$).  
The CP-conserving MSSM contains two $SU(2)$ Higgs doublets yielding five physical particles - four CP-even scalars ($h^0,\;H^0, \;H^- \;$ and~$H^+$) and one CP-odd
scalar ($A^0$)~\cite{mssmhiggs1,mssmhiggs2,mssmhiggs3}.  Here, $h^0$
is the lighter of the two neutral scalars.  
%At tree level, $m_h$ is the sole parameter in the Higgs sector of the SM, 
In the MSSM there are two parameters at tree level 
%and we may choose them 
that are conventionally selected to be $\tan{\beta}$ and $m_{A^0}$.  The parameter $\tan{\beta}$ is the ratio of the vacuum expectation values of the two Higgs doublets, and $m_{A^0}$ is the mass of the pseudoscalar Higgs particle.  
When $\tan{\beta}$ is large, the cross section for direct production of
Higgs bosons through gluon fusion becomes enhanced, making that
an appealing region for searches at the Tevatron.
The coupling strength of the $A^0$ boson to down-type
fermions with mass $m_f$ is proportional to $m_f\times\tan\beta$, hence couplings to tau leptons are enhanced.
The couplings of the $h_0$
are similarly enhanced for many possible models.
%In the MSSM, the coupling strength of $h^0$ or $A^0$ to a fermion is proportional to
%square of the product $m_f \times \tan{\beta}$
%$(m_f \times \tan{\beta})^2$, where $m_f$ is the fermion's mass; couplings to Tau leptons are enhanced. 

%At this writing, 

No fundamental scalar particle has yet been observed in any experiment.
%Indirect measurements of electroweak observables limit $m_h$ in the SM to be in the range $98^{+51}_{-35}$ GeV at 95\% confidence level~\cite{smhiggs_indirect}.  The combined data from the four experiments at the LEP $e^+e^-$ collider (ALEPH, L3, OPAL and DELPHI) have put the most stringent lower limit on $m_h$ in the SM at 114.4 GeV/$c^2$ at 95\% confidence level~\cite{smhiggs_direct}.  
The four experiments at LEP have 
each performed a search for $h^0/A^0$ produced in the
process: $e^+ e^- \rightarrow h^0 Z$ and $e^+ e^- \rightarrow h^0 A^0$.  
The combined results of four experiments have 
constrained the theory, excluding $m_{A^0}<91.9$~GeV/$c^2$, 
$m_{h^0}<91.0$~GeV/$c^2$ and $0.5<\tan{\beta}<2.4$ at 95\% 
confidence level \cite{lep_mssmhiggs}.
Another search for $h^0/A^0$ produced in association with two 
bottom ($b$) quarks and decaying to two $b$ quarks was performed 
at CDF earlier\cite{valls}.  
This previous search was sensitive to the high-$\tan{\beta}$ region, 
excluding $\tan{\beta}>50$ for $m_{A^0}=100$ GeV/$c^2$.

This paper presents the results of a search for 
supersymmetric (SUSY) %Supersymmetric 
Higgs bosons directly produced in 
proton-antiproton collisions at a center-of-mass energy of 
1.8 TeV using the $86.3 \pm 3.5$ pb$^{-1}$ of 
data recorded by the Collider Detector 
at Fermilab (CDF) during the 1994$-$1995 data taking period of the 
Tevatron (Run 1b).
Although the branching ratio to b quarks would be largest, 
that decay mode would be dominated by QCD background, so
we search for Higgs bosons that have decayed to two tau ($\tau$) leptons. 
Events are selected inclusively by requiring an electron from $\tau\rightarrow e\nu_{e}\nu_{\tau}$ and a hadronically decaying tau ($\tau_{h}$) lepton. 
This semi-leptonic mode was chosen for this search as a trade-off
between the distinctive electron signature in a QCD environment
and the high branching ratio of hadronic tau decays.

% The advantage of requiring that
%one tau decays to an electron
%is that its signature is distinct from the QCD backgrounds which
%are prevalent at hadron colliders.  We require the second tau decays 
%hadronically in order to take advantage of the high branching ratio
%for that channel; taus decay hadronically 65\% of the time.
%While the cross section for direct production of $h^0/A^0$ is 
%expected to dominate the cross section for the associated 
%production process sought in the previous analysis, a search 
%for directly produced $h^0/A^0$ particles decaying to $b$ quarks 
%would be dominated by QCD background. 

%On the other hand, hadronic decays claim a
%larger branching ratio, so if one is able to separate them from
%QCD jets, then there is much to be gained in the expected rate
%of detection. 

This is the first time that a search for Higgs bosons 
has been carried out in the 
%this 
di-tau decay mode using data from a hadron collider.  
In Run 1, CDF published other analysis with taus in
the final state~\cite{stops,franklin}.
%CDF has previously published a search for
%scalar top quarks in R-parity violating decay modes 
%with two taus in the final state~\cite{stops}. 
We also demonstrate for the first time from such data the 
feasibility of a technique to reconstruct the full mass of 
a candidate di-tau system, which is only possible when
the tau candidates are not back-to-back in the
plane transverse to the beam.  

%\bibitem{smhiggs_indirect} ``The Review of Particle Physics,'' K. Hagiwara et al., Phys. Rev. D 66, 010001 (2002).
%\bibitem{smhiggs_direct} ALEPH, DELPHI, L3 and OPAL Collaborations, The LEP Working Group for Higgs Boson Searches, ``Search for the Standard Model Higgs Boson at LEP'' Submitted to Phys.Rev.Lett.

The sample that passes the final selection cuts
is dominated by $Z\rightarrow \tau\tau$ events.
There is no evidence for a signal, 
so we report a limit on a set of MSSM models in a region of parameter
space where $\tan{\beta}$ is large because this is where the best
sensitivity is achieved. 
Since at the Tevatron most directly produced Higgs bosons would
be back-to-back, the acceptance is small in the region where the mass
reconstruction is a discriminating variable.  
Therefore,
limits are reported based on a counting experiment using events from
the full sample.  
Then, from a subset of the events where the
tau candidates are not back-to-back in the transverse plane 
we extract a mass distribution
and perform a binned likelihood 
to demonstrate the capability of that technique.

%Figure~\ref{feynman} depicts the Feynman diagrams for each process and Figure~\ref{cross section} the theoretical cross section for each.

%\Red{[...]}
%This paper is layed out as follows.
%The remainder of this section includes an overview of the analysis, 
%including a description of the detector with an emphasis on detector sub-systems critical to this analysis, as well as a description of the important issues related to the MSSM and event generation.  
%In Sec. III, we describe
%the event selection.
%In Sec. IV, we estimate the rate of backgrounds, both
%physics backgrounds and those
%that include fake hadronic taus.
%In Sec. V, the 
%systematics are quantified.
%In Sec. VI, the results are presented.  
%In Sec. VII, conclusions are drawn.

\section{The CDF Detector \label{section-detector}}
This section briefly describes the Run 1 CDF detector with an emphasis on the sub-detectors
important to this analysis.  The CDF detector is described in detail elsewhere~\cite{cdf1,cdf2,cdf3,cdf4}.

%The CDF detector measures
%the products of proton-antiproton collisions produced by
%the Tevatron.
%The Run 1 CDF detector was located at one of the six collision points along the Tevatron ring.    
%The nominal collision point 
%defines the origin of the CDF coordinate
%system, chosen to be right-handed.  
%The direction of positive $y$ was ``up,'' 
%so $x$ was on the plane of the Tevatron, pointing radially outward 
%from the center of the machine.  

CDF used a cylindrical coordinate system with the $z$ axis along the
proton beam direction.
The polar angle ($\theta$) was reported 
with respect to the $z$ axis. 
Pseudorapidity ($\eta$)
was defined as $-\ln [\tan (\theta/2)]$.
Detector pseudorapidity ($\eta_d$) was the same quantity with dependence on 
vertex position removed.  
The azimuthal angle ($\phi$) was measured relative to the positive 
$x$ direction.

The CDF electromagnetic and hadronic calorimeters were arranged in 
a projective tower geometry, as well as charged particle tracking chambers.
The tracking chambers were immersed in a 1.4 T magnetic field
oriented along the proton beam direction provided by a 3~m diameter,
5~m long superconducting solenoid magnet coil.  

%Drift chambers outside
%the hadron calorimeters for muon detection cover the region $|\eta|<1.0$.

In the central region covering $|\eta| < 1.1$, the electromagnetic (CEM) 
and hadron (CHA, WHA) calorimeters were made of absorber sheets interspersed
with scintillator.  Plastic light guides brought the light up to
two phototubes per EM tower.  The towers were constructed in 48
wedges, each consisting of 10 towers in $\eta$ by one tower in $\phi$.
The measured energy resolution for
the CEM and CHA were $\sigma(E)/E = 13.7\%/\sqrt{E_{T}} \oplus 2\% $
and $\sigma(E)/E = 50\%/\sqrt{E_{T}} \oplus 3\% $, respectively.
 
The central EM strip chambers (CES) 
were proportional strip and wire chambers located 
six radiation lengths deep in the CEM 
(radial distance $r=184.15$ cm), where the lateral size of
the electromagnetic shower was 
expected to be maximal~\cite{ces1,ces2,ces3,ces4,ces5}.  
%Figure~\ref{fig:CES} shows a segment of the CES.
It measured the position of electromagnetic showers
in the plane perpendicular to the radial direction with a resolution of
%$\lsim$ 1 cm 
2 mm
in each dimension~\cite{ces1,ces3}.
In each half of the detector (east and west), and for each
$15^{\circ}$ section in $\phi$,  the CES was subdivided into
two regions in $z$, with
128 cathode strips separated by $\approx$ 2 cm 
measuring the shower positions along the $z$ direction
with a gap within 6.2~cm of the $z=0$ plane.  
In each such region, 
64 anode wires (ganged in pairs) with a 1.45 cm pitch 
provided a measurement in $\phi$.  
%\begin{figure}
%\scalebox{0.52}{\includegraphics{ces.eps}}
%\caption{\label{fig:CES}Schematic of the CES.}
%\end{figure}

A three-component tracking 
system 
measured charged particle trajectories,
consisting of the silicon vertex detector (SVX$^\prime$), the 
time projection chamber (VTX) and the central tracking chamber (CTC).
The SVX$^\prime$~\cite{svxprime} consisted of four concentric silicon layers
sitting at radii between 2.36 cm and 7.87 cm and providing
$r-\phi$ tracking information only.
The VTX was positioned just beyond the
SVX$^\prime$ in radius and measured the 
position of the collision point along the beam 
for each event to a resolution of 2 mm.  

Beyond the VTX (radially) was the CTC, a cylindrical drift chamber 3.2~m long
in the $z$ direction, with its inner (outer) radius at 0.3 (1.3)~m.
%In the CTC, t
The sense wires were arranged into 84 layers divided into
9 ``super-layers.''  Five of the super-layers (axial) contained cells with 12 sense wires that ran
parallel to the beam and provided measurements in $r-\phi$.  The
remaining four super-layers (stereo) sat between the axial layers
in radius, contained 6 sense wires per cell, and 
were rotated in the $r-z$ projection
by 2.5$^{\circ}$ with respect to the beam to provide
measurements in $r-z$.  
The transverse momentum resolution of the CTC was 
$\delta p_{T}/p_{T} \lesssim  0.002 \text{ GeV/}c^{-1} \times p_{T}$
and when combined with the SVX$^\prime$ tracking information when
available, the resolution
was
$\delta p_{T}/p_{T}\lesssim 0.001 \text{ GeV/}c^{-1} \times p_{T}$.

\section{Monte Carlo Simulation \label{section-eventgen}}

The {\sc Pythia 6.203}~\cite{Pythia} event generator is used to simulate signal events and 
backgrounds other than fakes. The Monte Carlo (MC) samples that are 
generated with the standard {\sc Pythia} 
package 
required modifications that are discussed in this section.
%need some modifications
%before the events are passed through a 
%parameterized detector simulation called QFL.
%This section reviews the modifications made.

\begin{figure}
\scalebox{0.42}{\includegraphics{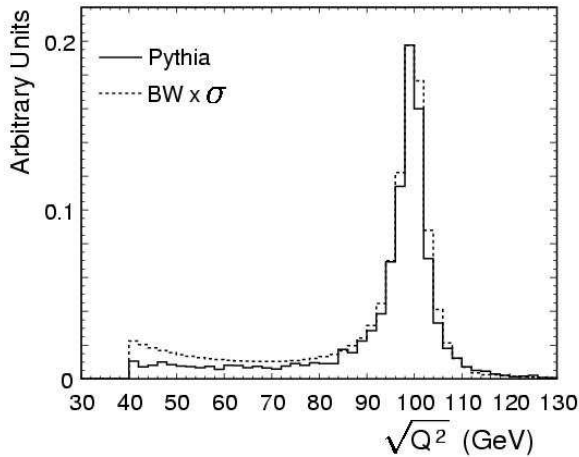}}
\caption{\label{fig:offshellhiggs} The figure shows the expected mass distribution of Higgs bosons produced directly at the Tevatron for $m_{A^0}~=~100$ GeV/c$^2$ and $\tan{\beta}~=~50$.  
We compare the output of {\sc Pythia 6.203}~\cite{Pythia} with that obtained from the method described in this paper:  folding in a simple relativistic
Breit-Wigner distribution with the 
mass-dependent cross section calculation.}
\end{figure}

A re-summed calculation of the 
cross sections for direct Higgs boson production at the Tevatron has
been performed using 
%Michael Spira~\cite{higlu}.  
the program \texttt{HIGLU}~\cite{higlu}, which 
allows a user to estimate cross sections as a function of mass,
$\tan{\beta}$
and other parameters.
For a given Higgs boson mass \texttt{HIGLU} gives
the \emph{on-shell} cross section only.  A 
%Supersymmetric 
SUSY Higgs boson 
produced at large $\tan{\beta}$ has a significant width and a tail
at low values of the center of mass energy of the parton collision
that produced the Higgs boson $\sqrt{Q^2}$  (see 
%Figure 
Fig.~\ref{fig:offshellhiggs}).  
Therefore, scaling the {\sc Pythia} differential distribution to the
\texttt{HIGLU} one would underestimate the cross section because
the off-shell events would not be accounted for.

To estimate the total cross section, we
first retrieve 
the on-shell cross section as a function of
mass in bins 1 GeV/$c^2$ wide using the \texttt{HIGLU} program.
%Next, we construct a simple Breit-Wigner shape:
%\begin{equation}
%\label{eq:breitwigner}
%\text{BW}(Q^2) = \frac{M_{A^0/h^0} \Gamma(Q^2)/\pi}{(Q^2-M_{A^0/h^0}^2)^2 + M_{A^0/h^0}^2 \Gamma(Q^2)^2} .
%\end{equation}
%The width is chosen to be proportional to $Q^2$:
%\begin{equation}
%\Gamma(Q^2) = \Gamma(M_{A^0/h^0}) \frac{Q^2}{M^2_{A^0/h^0}}.
%\end{equation}
%$\Gamma(M_{A^0/h^0})$ is taken to be the Higgs boson decay width from {\sc Pythia} 
%at the parameter space point of interest.
Then, the mass-dependent cross section is folded into a relativistic
Breit-Wigner shape, with the width proportional to $Q^2$,  
%For a given bin in $Q^2$, the cross section
%is given by:
%\begin{equation}
%\label{eq:foldin}
%\frac{1}{\pi}\cdot \sigma_{\text{HIGLU}}(Q_0^2) \cdot \left( \arctan{\frac{Q_+^2-M_{A^0/h^0}^2}{M_{A^0/h^0} \Gamma(Q_0^2)}} - \arctan{\frac{Q_-^2-M_{A^0/h^0}^2}{M_{A^0/h^0} \Gamma(Q_0^2)}}	\right)
%\end{equation}
%where $Q_+$ ($Q_-$) is the upper (lower) edge of the mass bin and
%$Q_0$ is the center of the mass bin. 
%$\Gamma(Q_0^2 = M_{A^0/h^0}^2)$ is the
%{\sc Pythia} width for a given Higgs boson mass and $\tan{\beta}$.
%We sum in (1~GeV)$^2$ bins 
from $Q^2=(40$~GeV$)^2$ to $Q^2=(200$~GeV$)^2$.
For $m_{A^0}=100$~GeV/$c^2$ at 
$\tan{\beta}=50$, the MSSM cross-section for $A^0+h^0$ production
is 122~pb. 

 The rate of Higgs boson production 
in the region of low $\sqrt{Q^2}$ is a source of significant
uncertainty for the analysis, particularly at high mass and high $\tan{\beta}$.  
The low tail seen in
%Figure
Fig.~\ref{fig:offshellhiggs} originates from a steeply falling
cross section folded in with an increase in parton luminosities
at small momentum transfer, folded in with a broad Higgs boson width.
When we use the 
method outlined above to obtain a $Q^2$ dependent cross section,
the size of the tail is bounded by those obtained when one generates
events at the same parameter space point 
with {\sc Pythia} and {\sc Isajet}~\cite{isajet}.
%In {\sc Pythia}, the Higgs boson coupling to lepton pairs is known to
%be incorrect in the region of high $\tan{\beta}$. 
We compare the result from the default {\sc Pythia} output 
with that where we use the method above to estimate the systematic
error from these low mass tails.
%We take the systematic error on the Higgs boson production cross section to be
%the percentage difference between the total efficiency of all of 
%our cuts when two different $Q^2$ dependent cross sections are used:
%first, the one from the 
%standard {\sc Pythia} output and second,
%the one
%obtained from Eqs.~\ref{eq:breitwigner} and ~\ref{eq:foldin}.
At $m_{A^0}=100$~GeV/c$^2$ and
$\tan{\beta}=50$, this uncertainty is $2\%$.  At higher mass and higher
$\tan{\beta}$, this systematic can become significant.  At $m_{A^0}=140$~GeV/c$^2$,
$\tan{\beta}=80$ the systematic is $30\%$.

\begin{figure}
\scalebox{0.44}{\includegraphics{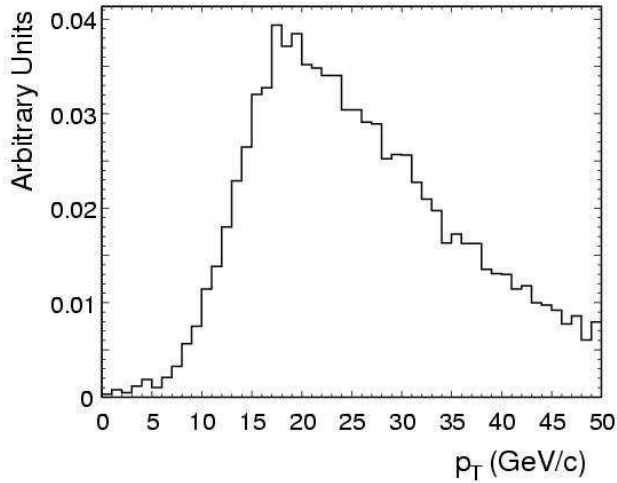}}
\caption{\label{fig:pt_dphicut} The $p_T$ of directly produced Higgs bosons with mass $m_A=100$ GeV after a
$\sin{\Delta\phi}>0.3$ cut has been imposed. }
\end{figure}

It is important to properly model the boson $p_T$ for 
Higgs boson and $Z$ boson events because it impacts
the relative rates of
back-to-back and
non-back-to-back events, and the di-tau mass resolution for the latter events.
%particularly in the non-back-to-back region
%where the mass technique is used.  
%The $p_T$ distribution is strongly correlated with the
%$\Delta\phi$ distribution, which affects 
%The mass resolution and
% are affected by the correlation between $p_T$
%and $\Delta\phi$, where $\Delta\phi$ is the angle between the
%taus in the transverse
%plane.
Figure~\ref{fig:pt_dphicut} shows the $p_{T}$ 
distribution of the $A^0$ ($m_{A^0}=100$ GeV/c$^2$) after the 
$\sin{\Delta \phi}>0.3$ cut is imposed, where $\Delta\phi$ is the
angle between the taus in the transverse plane.  
We see that this cut is 
approximately equivalent to a cut of $p_{T}>15$ GeV/c imposed on the 
parent Higgs bosons.

\begin{figure}
\begin{center}
\scalebox{0.3}{\includegraphics{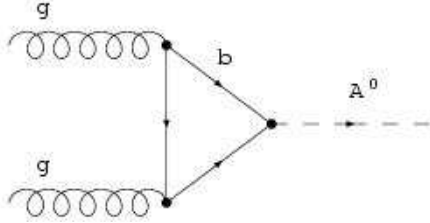}}
\caption{\label{fig:feynman}Feynman diagram of the
direct production process. In the high-$\tan{\beta}$ region of 
parameter space, the gluon fusion process occurs predominantly
through a triangular bottom quark loop. }
\end{center}
\end{figure}

In the high-$\tan{\beta}$ region of parameter space 
probed in this analysis, the direct-production 
process occurs predominantly through 
gluon fusion via a 
triangular bottom quark loop 
as in Fig.~\ref{fig:feynman} 
(the
direct production of SM Higgs bosons proceeds %Standard Model 
predominantly through a
$t$ %top 
quark loop).  
In the default {\sc Pythia}, the effect of the lighter quark mass
in the fermion loop on the
$p_T$ of Higgs bosons is not taken into account.  To correct for this,
we use~\cite{ptcalc} which 
performs a perturbative calculation for the differential 
cross section $d\sigma/dp_T$ to order $\alpha_s^{3}$ with a variable quark mass
in the triangular loop.  The perturbative calculation is valid in the
region $p_T^{Higgs}\gsim15$ GeV/c.  As has been pointed out, 
this cut is nearly equivalent to
the requirement $\sin{\Delta\phi}>0.3$.
 We force agreement between 
the Higgs boson $p_{T}$ distributions in the region
$p_{T}^{Higgs}>15$ GeV/c and the result of the program with a 
5 GeV/c$^2$ $b$ quark in the fermion loop so that
the resulting MC sample will have 
the proper efficiency for the $\sin{\Delta \phi}$ cut.
To do this, we use
a reweighting method known as the
 acceptance rejection MC %Monte-Carlo 
method~\cite{PDG}.    

The $p_T$ distributions for the $Z$ bosons from $Z/\gamma^*\rightarrow~e~e$
events has been measured by CDF 
in the region 66~GeV/c$^2<M_{ee}<$116~GeV/c$^2$~\cite{zpt}.  
$Z$ bosons from {\sc Pythia} tend to have a lower
average $p_T$ than the measured value.  Therefore, 
we reweight
both the $Z/\gamma^*~\rightarrow~\tau~\tau$ and 
$Z/\gamma^* \rightarrow ee$ samples 
to force
agreement between the MC events and the measured spectrum.  
Only events that lie in the measured region
are subject to rejection.
This correction makes a significant difference in the relative rates
of back-to-back and non-back-to-back events.  Before the correction,
20\% of the $Z$ events from {\sc Pythia} are in the non-back-to-back region 
(here, defined by $p_T^Z>15$ GeV/c).  After the correction, 26\%
of the events are in the non-back-to-back region.

CDF measured the cross section for $Z/\gamma^* \rightarrow e e$ 
in the same mass window as the $p_T$ distributions above, 
and reports this cross section to be 248~$\pm$~11~pb~\cite{zpt}.
Assuming universality, we
scale the generated $Z \rightarrow \tau \tau$ sample to this
cross section in the measured range.

Polarization effects, which impact the tau lepton kinematics, 
are properly taken into account with Tauola
~\cite{tauola}.
Simulated Higgs boson events tend to produce one tau
with hard (high-$p_T$) visible decay products and one with soft
visible decay products.  Taus produced from $Z$ bosons
are either both hard or both soft.

\section{Event Selection and Efficiencies \label{section-cuts}}

The search is performed in a data sample of events collected using
a high-$E_T$ electron trigger.
  In the following, we first describe the 
trigger system that an event must pass to enter
this data sample, followed by a description of the electron and  tau 
identification, and finally the event selection and mass reconstruction 
made at the analysis level.

\subsection{Trigger \label{sec:trigger}}

 A three-level trigger system was used to select events for \
data storage~\cite{l1l2trigger} and the 
energy-dependent efficiency of these triggers has been 
measured~\cite{Zee_eff}.
At Level 1 (L1), the 
trigger requires at least one trigger tower
with $E_T>8$ GeV in the CEM and the efficiency was measured to be 100\% at the
$10^{-3}$ level.
At Level 2 (L2), the trigger requires one
calorimeter cluster with $E_T~>~16$~GeV and the ratio of
hadronic energy to electromagnetic energy ($E_{HAD}/E_{EM}$) to be
less than 0.125.
This cluster must be close in azimuthal angle to a track reconstructed in the trigger with
$p_T>12$ GeV/c.  
The triggered sample
contains 128,761 events.
The efficiency of the L2 trigger for a good quality electron with
$E_T>20$ GeV
is $91\pm2$\%, measured from a $Z \rightarrow ee$ control sample.
Signal events would pass this trigger approximately
20\% of the time.  

\subsection{Electron Identification \label{sec:electid}}

To improve the purity of the data,
further cuts (listed in Table~\ref{tab:offlinecuts1})
are made on the candidate electron. 
\begin{table}
\begin{center}
\begin{tabular}{c} \hline \hline
$E_T>20$ GeV \\
$p_T>13$ GeV/c \\
$E/p<1.5$ \\  	
$Lshr< 0.2$ \\
$E_{HAD}/E_{EM}< 0.05$ 	\\		
$|\Delta x|<1.5$ cm  	\\
$|\Delta z|<3.0$ cm  	\\
$\chi^2_{strip}<10.0$  	\\ 
$|z_{\mathrm{VTX}}-z_e|<5.0$ cm  \\
$|z_{\mathrm{VTX}}|<60.0$ cm \\
Conversion Rejection \\
Fiducial cuts on the electron \\ \hline \hline
\end{tabular}
\caption{\sl{\label{tab:offlinecuts1}Cuts made offline to select the data sample used for
the search.  The electron identification variables are defined in Sec.~\ref{sec:electid}.
}}
\end{center}
\end{table}
%\begin{itemize}
%\item{$E_T>20$ GeV}
%\item{$p_T>13$ GeV}
%\item{$E/p<1.5$}
%\item{Lshr$_2$ $< 0.2$} 	
%\item{Lshr$_3$ $< 0.2$} 
%\item{$E_{had}/E_{em}(2x3)< 0.125$} 
%\item{$E_{had}/E_{em}(3x3)< 0.05$} 			
%\item{$|\delta x|<1.5$ cm}  	
%\item{$|\delta z|<3.0$ cm}  		
%\item{$\chi^2_{strip}<10.0$}  	 
%\item{$|Z_v-Z_0^e|<5.0$ cm (Vertex Class $\ge 10$)}
%\item{$|Z_v|<60.0$ cm}
%\item{Fiducial cuts on the electron}
%\end{itemize}
These cuts are described in more detail in~\cite{johnwahl};
we briefly summarize them here.
We use $Z \rightarrow ee$ control samples to quantify the degree of agreement
between the simulation and the data.

A candidate electron is first identified as a calorimeter cluster in the CEM.
An electron cluster in the calorimeter
is formed by merging seed towers (required to have
$E_T>3$ GeV) with neighboring towers in $\eta$ 
($E_T>0.1$ GeV required).  To be called an EM cluster, 
$E_T>5$ GeV is required.  Corrections are made to the 
energy of an EM cluster to compensate for variable response across 
each tower, tower-to-tower gain variations and time-dependent
effects.  A global correction to the 
energy scale is also imposed~\cite{saltzberg}.  
These corrections are typically at the level of a few percent.

We require an electron candidate to have a CTC track 
pointing 
%to a tower in its EM cluster.  
to an EM cluster.
The highest $p_T$
track pointing to the cluster is the ``electron track,''
and is required to satisfy $p_T>13$ GeV/c.
The measured direction of the track momentum 
sets the direction of the electron
candidate.

We require
that the ratio of the energy deposited in the EM calorimeter
to the momentum of the electron track is not too large ($E/p<1.5$).
Also, the lateral profile of the EM shower left by the electron tau candidate
is required to be consistent with electron shower profiles as measured
in test beam data ($Lshr<0.2$). 
The ratio of energy measured in the
CHA to energy measured in the CEM ($E_{HAD}/E_{EM}$) is expected
to be small for an electron from a tau decay; 
we require $E_{HAD}/E_{EM}<0.05$. 
In addition, the electron track 
projected to the plane of the CES must be close to a CES
shower position:  $|\Delta x|<1.5$ cm and 
$|\Delta z|<3.0$ cm.  This reduces the background from charged pions
produced in the neighborhood of neutral pions.  
We also confirm that the profile of the pulse heights produced
by the electron shower across CES strips 
is consistent with electron test beam data:  $\chi^2_{strip}<10.0$.
The electron track is required to be consistent with a
vertex 
%($|z_e-z_{\mathrm{VTX}}|<5$ cm) 
that lies within 60 cm of the nominal collision point.  
Standard CDF fiducial cuts are made on the electron to ensure
that the particle arrived at an instrumented region of the calorimeter 
with good response.  
We reject electron candidates which are consistent with having
originated from a photon that converted to an electron-positron pair
in the detector by removing candidates 
that leave a low-occupancy track in the VTX or that have 
a nearby  opposite-sign track.
The opposite-sign track must be within $90^{\circ}$ in $\phi$, 
separated in $r-\phi$ from the electron track
by no more than 0.3 cm measured at the point where the tracks are
parallel,
satisfying $|\Delta \cot{\theta}|<0.06$ where 
$\Delta \cot{\theta}$ is the difference between the values of
$\cot{\theta}$ for the two tracks.
The electron tau candidate must be isolated in the calorimeter
and in the tracking system.  
In the calorimeter, we use the standard CDF isolation variable defined by:
\begin{equation}
	R_{\mathrm{iso}} = \frac{E^{\mathrm{cone}}-E^{\mathrm{cluster}}}{E^{\mathrm{cluster}}}
\end{equation}
where $E^{cone}$ is the energy deposited in the calorimeter in a 
cone of $\Delta R~=~\sqrt{(\Delta\phi)^2+(\Delta\eta)^2}~=~0.4$
%$\Delta~R~=~0.4$ ($\Delta R~=~\sqrt{(\Delta\phi)^2+(\Delta\eta)^2}$)
around the electron tau candidate and $E^{\mathrm{cluster}}$ is the energy of the
EM cluster.  We require $R_{\mathrm{iso}}<0.1$.
We define $N_{iso}$ to be the number of tracks with $p_T>1$ GeV/c 
within $\Delta R<0.524$ of the EM cluster.
This cone size is 30$^\circ$, chosen
to be the same as the outer radius of the isolation annulus 
used for identifying hadronic taus, described below.
  A track must
originate within 5 cm of the electron track along the beam line
to be counted in the isolation cone.
We require that $N_{iso}=0$.

We require at least one EM cluster in the event with $E_T>20$ GeV 
passing the electron identification requirements just described. 
We refer to this as the
\emph{electron tau ($\tau_{e}$) candidate}.  
If there is more than one $\tau_{e}$ candidate in the event, 
we select the candidate with highest $E_T$.

We correct for inadequacies in simulation of electron identification
variables by applying a scale factor of 
 $0.869 \pm 0.016$.  The scale factor was determined using
$Z/\gamma^* \rightarrow ee$ events and we have determined that is
not $p_T$ dependent.

Since the vertex position affects the acceptance, we 
reweight the events to
force agreement
between the distribution of primary 
vertex positions from the simulation and that measured from
the Run 1b data sample.

\subsection{Hadronic Tau Identification\label{sec:tauid}}

\begin{table}
\begin{center}
\begin{tabular}{c} \hline \hline
$E_T>10$ GeV (jet)\\
$p_T>10$ GeV/c (track)\\ 	
$N^{\mathrm{trks}}_{\mathrm{cone}}=1$ or $3$\\
$N^{\mathrm{trks}}_{\mathrm{ann}}=0$\\
Fiducial requirements \\ 
$|z_{\mathrm{VTX}}-z_{\tau_h}|<5.0$ cm  \\
$|z_{\tau_h}|<60.0$ cm \\ 
$N^{\mathrm{ces}}_{\mathrm{cone}}<3$ \\
$N^{\mathrm{ces}}_{\mathrm{ann}}=0$ \\
$m_{\tau}<2.0$ GeV/c$^2$ \\ 
$\xi>0.15$ \\
$I_{\phi-\phi} < 0.1$ and $I_{\eta-\eta} < 0.1$ \\ \hline \hline
\end{tabular}
\caption{\sl{\label{tab:offlinecuts2}Cuts made offline for tau identification.  These variables are defined in Sec.~\ref{sec:tauid}.
}}
\end{center}
\end{table}

% SUGGESTION: Can we put this block in the reference?

%The lifetime of a tau at rest is 290.6 $\pm$ 1.1 fs~\cite{PDG},
%so we do not observe taus directly but instead we observe their 
%decay products.  A tau may decay to other leptons or hadrons. 
%85.35~$\pm$~0.07\% of those decays are ``one-prong'' (the 
%decay products include only one charged particle), 
%the hadronic one-prong claiming 49.5\% of the total branching ratio.  
%Three-prong decays occur 15.20~$\pm$~0.07\% of the time.  
%One-prong hadronic decays contain 
%a neutral pion nearly half the time but rarely contain more than two.  
%Three-prong decays
%contain one neutral pion at most.  
%\Red{[]} %We call a hadronic tau candidate the \emph{second tau candidate}, or $\tau_h$.  

For taus coming from Higgs boson decays, the hadronic tau decay products will
be collimated, with an angular deviation from the 
direction of the tau parent of no more than 
$\Delta \phi~\lsim~m_{\tau}/E_{\tau}$ which is $\sim 10^{\circ}$ for a
typical $E_{\tau} \sim 10$ GeV.  
In nearly all cases, 
tau decay products will include 1 or 3 charged tracks and 
$\le 2$ neutral pions, each decaying to two photons (all other
decay modes have branching fractions of less than half of a 
percent). 

The cuts used to select a hadronic tau are listed in
Table~\ref{tab:offlinecuts2}. The $\tau_h$ %tau 
identification cuts used here are based on those outlined 
in~\cite{previoustau}.
The main differences are noted in what follows.
The search for a 
$\tau_h$ %hadronic tau 
begins with identifying a stiff track
associated with a jet cluster.  We require a track with $p_T>10$ GeV/c 
within $\Delta R<0.4$
 of a jet cluster with $E_T>10$ GeV. 
The calorimeter cluster size is $\Delta R=0.4$.
The track with
the highest $p_T$ satisfying this requirement is called the \emph{tau seed}.
The $E_T$ cut is approximately
75\% efficient for signal.
%while keeping 70\% of the  
%inclusive electron sample 
%(where the tau candidate is predominantly a fake
%from a QCD jet).
The $p_T$ cut is
approximately 65\% efficient for signal while rejecting 80\% of
QCD jets (after the $E_T$ cut has already been imposed).

We make stringent fiducial requirements on the tau seed
to ensure that the tau candidate's energy is well measured.  
%The track must enter the central calorimeter and it must stay 
%within the tracking volume of the CTC up to its outer radius.  
The track must be fully contained in the CTC and
 must pass additional fiducial cuts similar to those
imposed on the electron candidate 
to ensure that it is not incident on an uninstrumented portion of the
calorimeter.  Also, to suppress fake track contamination,
we require 0.5 GeV in 
the tower to which the track points.
The seed track also must be within 5 cm of the same primary vertex
($|z_{\mathrm{VTX}}-z_{\tau_h}|<5.0$ cm) as the electron track.

In the neighborhood of the tau seed,  
two isolation 
regions are considered separately with different requirements
made in each region: 
the $\Delta R<0.175$ cone and the 
$0.175~<~\Delta R~<~0.524$ annulus.
%\Red{[]} % (A cone size of 0.175 (0.524) radians corresponds to $10^{\circ}$ ($30^{\circ}$)).
In either isolation region, a track is a \emph{shoulder} 
track if it is a good quality track with
$p_T>1$ GeV/c.
The seed track is included in the track counting in the $R<0.175$ cone.

For a true hadronically decaying tau, 
the number of tracks in the $\Delta R<0.175$ cone 
($N^{\mathrm{trks}}_{\mathrm{cone}}$)
is usually 1 or 3, so  
we require $N^{\mathrm{trks}}_{\mathrm{cone}}<4$.  
%We then sum the 
%charges of the tracks in the cone, which we call $Q$, and
%require $|Q|$=1.  We also require that $Q$ is opposite in sign from
%the electron charge.
We additionally require the sum of the charges of the tracks in the cone
to be $\pm 1$, and opposite to the charge of the electron.
We expect the number of tracks in the annulus $0.175<R<0.524$
around the tau seed ($N^{\mathrm{trks}}_{\mathrm{ann}}$) to vanish in signal events, 
so we require $N^{\mathrm{trks}}_{\mathrm{ann}}~=~0$.
These tracking isolation 
cuts retain approximately 80\% of signal events and reject 70-80\% of QCD jets.

The CES clustering algorithm used is the same as
the one used in previous CDF analyses~\cite{previoustau},
with some modifications that improve tau purity and fake rejection.  
In particular,
CES clusters were formed at a larger distance in $r-\phi$ 
from the seed track so that this information may be used for
fake rejection.  Also, a $\chi^{2}$ requirement that was used in previous CDF analyses to 
ensure the consistency of the CES cluster profile with electron test
beam data was removed here to improve efficiency without a 
significant sacrifice in purity.
%We also add some fiducial requirements that 
%improve the agreement between the data and 
%simulated MC.

The algorithm
forms CES clusters in the
$\Delta R<0.6$ cone around the seed track
by taking the highest energy strips (wires) in each
calorimeter tower, in descending order, calling them seeds, 
and merging them with their nearest neighbors
to form clusters
4 strips (6 wires) wide.
To be a seed for a cluster, a strip (wire) must show a 
pulse height that surpasses 0.4 (0.5)~GeV.
 The cluster position 
is defined as
the position of the center strip or wire in the cluster.
Pulse heights from strips and wires were corrected for $\eta-$ and
$\phi-$dependent effects, measured 
from test-beam data.  
The energy of a CES cluster in a tower is 
the CEM energy of that tower, weighted by the energy 
deposited in the CES by that cluster compared to the
energy deposited by all CES clusters in the tower. 
The predicted response in
the CEM for a charged pion is subtracted from the
energy in the tower impacted by the seed track.
Wire clusters in the $r-\phi$ view are 
matched with strip clusters in the $z$ view 
with similar energies
and merged into new clusters.
The cluster position must not be consistent with coming from the
seed track.  It is rejected if 
$|\eta_d^{\mathrm{seed}}-\eta_d^{\mathrm{cluster}}|<0.03$ and either 
$|\phi^{\mathrm{seed}}-\phi^{\mathrm{cluster}}|<0.01$ or 
$|\phi^{tower\hspace{0.05in} center}-\phi^{\mathrm{cluster}}|<0.01$
(the latter requirement is because
the cluster position is assigned
to the center of the tower in $\phi$ when no wire information
is available).

A CES cluster must satisfy $E_T>1$ GeV
to be counted as a \emph{shoulder cluster} in either of the 
isolation regions.
We call the number of CES clusters found in the isolation cone (annulus)
$N^{\mathrm{ces}}_{\mathrm{cone}}$ ($N^{\mathrm{ces}}_{\mathrm{ann}}$) and require 
$N^{\mathrm{ces}}_{\mathrm{cone}}<3$ and $N^{\mathrm{ces}}_{\mathrm{ann}}=0$.
In addition,  
the CES cluster energies in the cone 
measured as 3-component vectors
are combined with the measured momenta of the tracks to compute 
a tau mass, $m_{\tau}$.  We require $m_{\tau}<2.0$ GeV/c$^2$.
The cuts on  $N^{\mathrm{ces}}_{\mathrm{cone}}$, $N^{\mathrm{ces}}_{\mathrm{ann}}$ and $m_{\tau}$
give a combined efficiency for signal of approximately 95\%.
These three cuts additionally reject 30-50\% of QCD jets.
%keep 50-70\% of QCD jets.

As described above, a tau candidate is found to be isolated
through a measurement of track and CES cluster multiplicities
in the neighborhood of a seed track.
We compare the isolation variables among
$Z\rightarrow ee$, $Z\rightarrow \mu\mu$ and $Z \rightarrow \tau\tau$ 
simulation MC samples, and in $Z\rightarrow ee$ and $Z\rightarrow \mu\mu$ 
data control samples.  In the $Z \rightarrow ee$ and
$Z \rightarrow \mu\mu$ samples, both lepton candidates
in the event mimic the tau seed in this analysis.
We find good agreement 
between
data and simulation, and between electron and muon samples,
in the efficiencies of the isolation cuts.
The isolation efficiencies from the 
simulated $Z \rightarrow \tau \tau$ MC sample also agrees with
the data samples in the annulus around the tau seed (where no particles
from tau decays are expected).  
%We find that a correction
%to the CES cluster energies (derived from the CEM energy as
%described above) needed a modest correction.  This correction
%has a negligible effect on this analysis since the CES cluster
%energy is only used for the $m_{\tau}$ cut, which is nearly
%100\% efficient for taus.

To reduce the impact of $Z/\gamma~\rightarrow~ee$ background on the 
sensitivity,
we require $\xi>0.15$ where
$\xi~=~E^{had}_T / \Sigma{p_T}$~\cite{leslie}.  
Here, $E^{had}_T$ is the hadronic energy
of the tau jet cluster 
%(uncorrected) 
and
$\Sigma{p_T}$ is the sum of the $p_T$ of
all tracks with $p_T>1$ GeV/c within the cone $\Delta R<0.175$ 
centered on the jet direction.  
%By comparing 
%$Z \rightarrow ee$ data control samples with MC samples we find that
%the simulated efficiency for this cut for events where
%the hadronic tau candidate is an electron needs a 
%corrective factor of $1.9 \pm 0.4$, but this background is
%very small so the corrective factor has little impact.

%which only needs to be applied
%to a fraction of an event as modeled in each of the  
%$Z\rightarrow ee$ and $Z\rightarrow \tau\tau$ MC samples,
%and even fewer in the $A^0/h^0\rightarrow \tau\tau$ samples.

Since the decay products of a hadronically decaying tau are highly collimated,
a tau is expected to leave a narrow cluster of energy in the calorimeter.
We cut on the $\phi-\phi$ 
and $\eta-\eta$ moments of the jet cluster associated with the hadronic 
tau candidate which are defined as follows:
\begin{equation}
I_{\phi-\phi} = \frac{\sum_{i}{E_T^{\hspace{0.05in}i} \cdot {(\phi_i-\phi_0)}^2}}{\sum{E_T^{\hspace{0.05in}i}}} 
\end{equation}
\begin{equation}
I_{\eta-\eta} = \frac{\sum_{i}{E_T^{\hspace{0.05in}i} \cdot {(\eta_i -\eta_0)}^2}}{\sum{E_T^{\hspace{0.05in}i}}}.
\end{equation}
The sum is over calorimeter towers in the jet cluster, 
and $\phi_0$ and $\eta_0$ are the
$E_T$ weighted center of the jet in the $\phi$ and $\eta$ directions.
We require $I_{\phi-\phi} < 0.1$ and $I_{\eta-\eta} < 0.1$.  
These
two cuts together reject approximately 30-45\% of QCD jets.

\begin{figure}
\scalebox{0.42}{\includegraphics{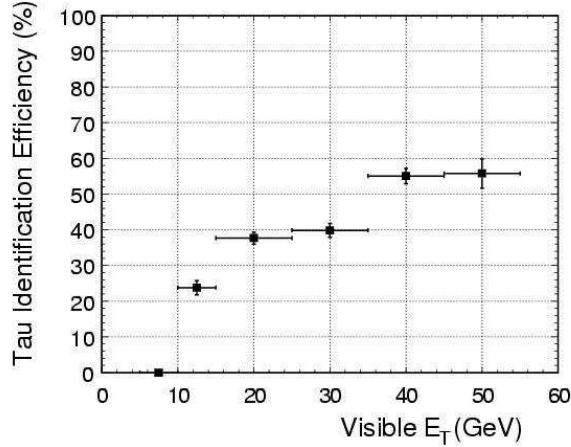}}
\caption{\label{fig:ztt_eff} Efficiencies of the tau identification cuts applied to 
hadronic taus from $Z \rightarrow \tau \tau$ events .}
\end{figure}

Figure~\ref{fig:ztt_eff} shows the efficiency of the hadronic tau 
identification cuts from the
simulation.  
We bin the efficiencies according to 
the visible transverse energy ($E_T$)
from the tau at the 
generator level.  
The
total efficiency plateaus near 55\% for $E_T\gsim 35$ GeV.
For fiducial hadronic taus with $E_T>10$ GeV from
$Z$ decays, the average efficiency of the
tau identification cuts is
40\%.  $A \rightarrow \tau \tau$
events at $m_A=100$ GeV and $\tan{\beta}=50$ are also 40\% efficient.

%We also require that the sum of the measured charges of the
%tracks in the 10$^\circ$ cone surrounding the hadronic tau candidate
%is opposite in sign from the electron tau candidate (the electron candidate).
%The efficiency of this cut is measured from data control samples as
%described in Sec.~\ref{chap:ANA2}.

\subsection{Additional Requirements\label{kinem}}

We make further cuts 
on the data sample to increase purity and further reject background.
We require at most one recoil jet with $E_T>15$ GeV in the event.
For further suppression of $Z\rightarrow ee$ background,
we reject any event with two electrons or one electron and one track
that have a reconstructed invariant 
mass $M_{ee}$ between 70 and 110 GeV/c$^2$, which
removes
99\% of $Z\rightarrow ee$ events
while retaining 90\% of signal events.
We require a separation of the tau candidates in the
transverse plane, $\Delta\phi>1.5$, where
$\Delta\phi$ is the azimuthal angle between 
$\tau_e$ and $\tau_h$.  This is nearly
100\% efficient for signal rejecting 20\% of non-tau backgrounds
as measured from a background-dominated data sample.

To take advantage of the mass reconstruction technique,
we divide the events into back-to-back and non-back-to-back
samples.
The full invariant mass of the di-tau system
can be estimated only when the tau candidates are not back-to-back,
as explained in more detail in Section IV E.  The tau candidates are 
called back-to-back when
$\sin\Delta\phi < 0.3$, where $\Delta\phi$ is the
azimuthal angle between $\tau_e$ and $\tau_h$.  
Note that $\sin\Delta\phi$ is the determinant
of the system of equations which determine the di-tau mass, so when
$\sin\Delta\phi \approx 0$, the solution is not unique.

%The events where the tau candidates are
%back-to-back in the transverse plane 
%are considered separately from events where the tau candidates are
%not back-to-back.
%We do this because when the tau candidates are not back-to-back,
%each tau's contribution to the event missing energy, and thus 
%the full mass of the di-tau system may be determined using the
%technique described in Sec.~\ref{sec:massreconstruction}.
%We say that the tau candidates are back-to-back 
%if they satisfy 
%$|\sin{\Delta\phi_{\tau_e,\tau_h}}|<0.3$ and $\Delta\phi>1.5$,
%whereas the non-back-to-back events satisfy 
%$|\sin{\Delta\phi_{\tau_e,\tau_h}}|>0.3$.
%This strange-looking variable, $\sin{\Delta\phi}$, is chosen because it
%is the determinant of the system of equations solved to determine the di-tau
%mass.  When $\sin{\Delta\phi_{\tau_e,\tau_h}}=0$, 
%the system cannot be solved.  
%$\sin{\Delta\phi}=0.3$ and $\Delta\phi>1.5$ corresponds to $\Delta\phi \approx 160^{\circ}$.

%We call the event-wide energy imbalance in the transverse plane
The missing transverse energy in the event, denoted 
$\vec{E_{T}}\hspace{-0.18in}/\hspace{0.1in}$,
is the opposite of the vector sum of the measured transverse
energies of the event.
%Theoretically, the
%missing transverse energy ($E_{T}\hspace{-0.18in}/\hspace{0.1in}$)
%is equal in
%magnitude to the sum of visible energy in the transverse plane for
%a particular event, and points in
%the opposite direction.  In practice, its
%$E_{T}\hspace{-0.18in}/\hspace{0.1in}$ depends on 
% jet energy corrections, electron energy corrections,
%etc. 
For the non-back-to-back events, we 
use the magnitude and direction of 
$\vec{E_{T}}\hspace{-0.18in}/\hspace{0.1in}$ to derive 
the di-tau mass.
First, we define corrected 
$E_{T}\hspace{-0.18in}/\hspace{0.1in}$ in the following way:
\begin{equation}
\vec{E_{T}}\hspace{-0.18in}/\hspace{0.1in}^{corr} = - \sum_{towers}{\vec{E_{T}}^i}  - \sum_{muons}{\vec{p_{T}}^j} - \Delta \vec{E_{T}}^{ele} - \sum_{jets}\Delta\vec{E_{T}}^i . 
\end{equation}
The first term on the right side of the equation is a sum of the 
the transverse component of the energy deposited the calorimeter towers.
The remaining terms improve the resolution by accounting
 for the momentum carried away by muons,
energy corrections applied to the electron candidate, and 
jet energy corrections.

%As described in Sec.~\ref{sec:massreconstruction}, 
%the di-tau mass may only be reconstructed when
%the taus are not back-to-back, and 
It is only necessary for the
simulation to correctly model $\vec{E_{T}}\hspace{-0.18in}/\hspace{0.1in}^{corr}$
 well in the region
$p_T^{A,h,Z}>15$ GeV/c, since that is where the mass reconstruction is utilized. 
%Using a
%$Z \rightarrow ee$ data control sample and 
%a $Z \rightarrow ee$ MC Sample, 
We confirm that the $\vec{E_{T}}\hspace{-0.18in}/\hspace{0.1in}^{corr}$
variable from the data is well modeled by the 
simulation  
using a sample of $Z\rightarrow ee$ events with $p_T^{Z}>15$ GeV/c.
%does model $\vec{E_{T}}\hspace{-0.18in}/\hspace{0.1in}^{corr}$ well in this region.

\subsection{Di-tau Mass Reconstruction}
\label{sec:massreconstruction}

Signal events contain neutrinos that escape CDF
undetected.  At hadron colliders, the resulting energy 
imbalance may only be determined in the transverse plane 
%because the
%center-of-momentum of the interaction is not known along the $z$ 
%direction
because the $z$-component of the total
momentum of the interaction is unknown.

Nonetheless, the energy of the neutrinos from each tau decay, and thus
the full mass of the di-tau system, may be deduced if
(i) the tau candidates are not back-to-back in the transverse plane and
(ii) the neutrino directions are assumed to be the same as their 
parent taus\cite{ptcalc,cms,atlas1,atlas2}.

The contributions to the total missing energy from
 the leptonic and hadronic decays, denoted
$E_l\hspace{-0.13in}/\hspace{0.1in}$ and $E_h\hspace{-0.16in}/\hspace{0.1in}$, 
are the solution to a system of two equations and two unknowns:
\begin{equation}
\label{eq:me1}
 E_{l}\hspace{-0.13in}/\hspace{0.1in} 
\sin{\theta_{l}}\cos{\phi_{l}} 
+ E_{h}\hspace{-0.16in}/\hspace{0.1in}\sin{\theta_{h}}\cos{\phi_{h}} 
= (E^{meas}\hspace{-0.38in}/\hspace{0.32in})_{x} 
\end{equation}
\begin{equation}
\label{eq:me2}
 E_{l}\hspace{-0.13in}/\hspace{0.1in}
\sin{\theta_{l}}\sin{\phi_{l}} 
+ E_{h}\hspace{-0.16in}/\hspace{0.1in}\sin{\theta_{h}}\sin{\phi_{h}} 
= (E^{meas}\hspace{-0.38in}/\hspace{0.32in})_{y}
\end{equation}
Here, $E^{meas}\hspace{-0.38in}/\hspace{0.32in}$ is the
missing energy measured for the event and
$\theta_{l,h}$ are the polar angles of the
taus and 
$\phi_{l,h}$ are the azimuthal angles of the taus.
The tau candidate directions are measured from the visible decay
products.

%Figure~\ref{fig:massdphi} highlights the reason for 
The reason for the first
of the two criteria for the mass technique outlined above is that
when the tau candidates are back-to-back in the transverse plane,
the reconstructed mass is not a good separating variable because
there are many high mass solutions.
%\begin{figure}
%\scalebox{0.42}{\includegraphics{mass_dphi_test.ps}}
%\caption{\label{fig:massdphi} For Higgs boson events with $m_A=100$ GeV and
%$\tan{\beta}=50$, this figure shows the reconstructed 
%di-tau mass plotted against the azimuthal
%angle between the tau candidates.  Here, only events with a nonzero
%solutions for $E_{l}\hspace{-0.13in}/\hspace{0.1in}$ and 
%$E_{h}\hspace{-0.16in}/\hspace{0.1in}$ appear on this plot.}
%\end{figure}
%We have chosen $|\sin{\Delta\phi}|>0.3$ to select non-back-to-back 
%events.  This cut is approximately 20\% efficient for $A^0/h^0 \rightarrow \tau\tau$
%and $Z \rightarrow \tau \tau$ events.  With a stiffer cut, the mass
%resolution would improve, as can be seen from Fig.~\ref{fig:massdphi},
%but the efficiency of the $|\sin{\Delta\phi}|$ cut 
%would be significantly 
%degraded.  
%The second of the criteria is necessary since we do not detect
%the directions of the neutrinos as they leave the detector.
%As mentioned above, the decay products of a tau in the events
%of interest are deflected by no more than $\sim10^{\circ}$.
% plot showing how this affects the mass reconstruction?

\begin{figure}
\scalebox{0.48}{\includegraphics[width=\textwidth]{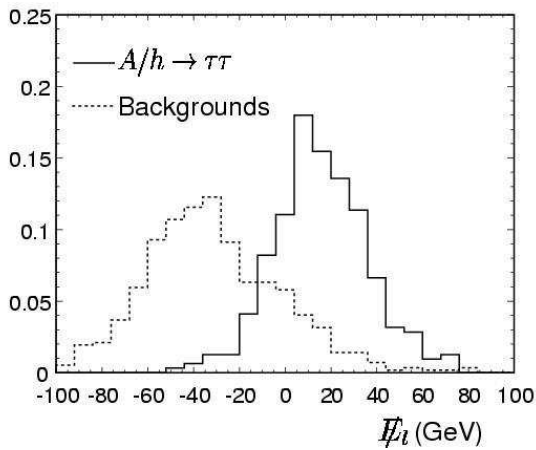}}
\caption{\label{fig:me1} $E_{l}\hspace{-0.13in}/\hspace{0.1in}$ from simulated
$A^0/h^0 \rightarrow \tau \tau$ events compared to a background-dominated
data sample.}
\end{figure}

%Due to the the criteria necessary for the mass technique and
%due to the resolution of the detector, 
Some events may give
negative solutions for $E_{l}\hspace{-0.13in}/\hspace{0.1in}$ and 
$E_{h}\hspace{-0.16in}/\hspace{0.1in}$.  
We require $E_{l,h}\hspace{-0.23in}/\hspace{0.12in}>0$ for the 
non-back-to-back
events, which reduces non-di-tau backgrounds in this region.
Figure~\ref{fig:me1} 
%Figure~6
shows 
$E_{l}\hspace{-0.13in}/\hspace{0.1in}$ from a background-dominated
sample of the data compared to simulated
$A^0/h^0 \rightarrow \tau \tau$ events.  The analogous distributions for
$E_{h}\hspace{-0.13in}/\hspace{0.1in}$ are similar.
The $E_{l,h}\hspace{-0.23in}/\hspace{0.12in}>0$ requirement 
is 55-60\% efficient for signal, increasing with mass, 
while reducing non-di-tau background by approximately a factor of 10.
When the di-tau mass is reconstructed as described next,
this cut also improves the mass resolution.

We calculate the total reconstructed mass of the di-tau system using
\begin{equation}
\begin{split}
m_{\tau\tau}^{2} &= m_{Z/h}^{2} = (p_{l} + p_{h})^{2} \\ 
%\>$ = p_{l}^{2}+p_{h}^{2}+2p_{l}\cdot p_{h}$ \\ 
%\>$ = 2m_{\tau}^{2}+2E_{l}E_{h}(1-\cos{\psi})$ \\ 
&= 2m_{\tau}^{2}+2
(E_{l}\hspace{-0.13in}/\hspace{0.1in}+E_{l}^{vis})
(E_{h}\hspace{-0.16in}/\hspace{0.1in}+E_{h}^{vis})(1-\cos{\psi}) \\
\end{split}
\end{equation}
where $p_{l}$ and $p_{h}$ are the 4-momenta of each tau, and
$E_{l}$ and $E_{h}$ represent the total energy of each tau.
$E_{l}^{vis}$ and $E_{h}^{vis}$ represent
the energy left by their
visible decay products.
The $\psi$ is the 3-dimensional angle between the two taus.
The missing energies $E_{l}\hspace{-0.13in}/\hspace{0.1in}$ and 
$E_{h}\hspace{-0.16in}/\hspace{0.1in}$ are the solutions to
Eqs.~\ref{eq:me1} and ~\ref{eq:me2}.

\begin{figure}
\scalebox{0.4}{\includegraphics{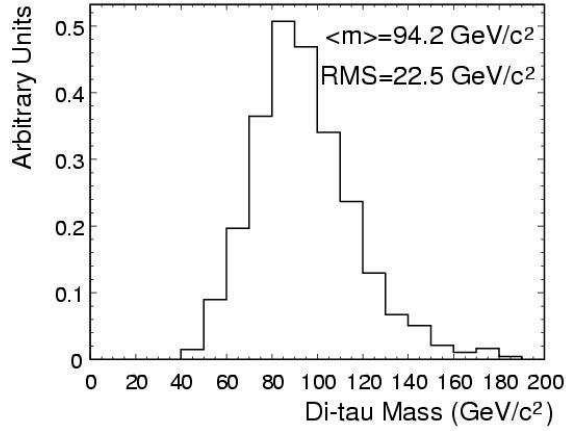}}
\caption{\label{fig:massplot_sim}Di-tau mass distribution as modeled by {\sc Pythia 6.203} and the CDF
detector simulation with parameters
$m_A=100$ GeV and $\tan{\beta}=50$.  
A $\sin(\Delta\phi)>0.3$ cut has been imposed.}
\end{figure}
\begin{figure}
\scalebox{0.4}{\includegraphics{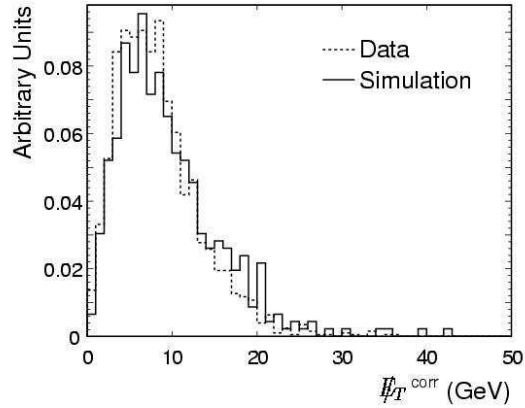}}
\caption{\label{fig:met_co_ptcut}Corrected missing energy as
measured from a $Z \rightarrow ee$ data control sample, only for events
where $p_T^Z>15$~GeV/c.}
\end{figure}
%Figure~\ref{fig:massplot_sim} 
Figure~\ref{fig:massplot_sim}
shows the di-tau mass distribution
reconstructed from signal events
for the parameter point $m_{A^0}=100$~GeV/c$^2$ and  $\tan{\beta}=50$,
 for which the $A^0$ or $h^0$ particle has an inherent width of 5.7 GeV/c$^2$.
The contribution from the calorimeter resolution may be qualitatively
seen from 
%Fig.~\ref{fig:met_co_ptcut} 
Fig.~\ref{fig:met_co_ptcut}, since $Z \rightarrow ee$
events would have a small $E_{T}\hspace{-0.18in}/\hspace{0.1in}$.  
The remaining contribution to the
width of the di-tau mass distribution
comes from
the approximations that are needed to implement the technique, listed above.  
%Notice that the mass reconstruction gives a mean that is low by approximately 6\%.
%\begin{figure}
%\scalebox{0.42}{\includegraphics{mass_dphi.eps}}
%\caption{\label{fig:massplot_sim} Reconstructed di-tau mass plotted against the azimuthal
%angle between the tau candidates.  Here, only events with a nonzero
%solutions for $E_{l}\hspace{-0.13in}/\hspace{0.1in}$ and 
%$E_{h}\hspace{-0.16in}/\hspace{0.1in}$ appear on this plot.}
%\end{figure}

\begin{table}
\begin{center}
\begin{tabular}{lcc} \hline \hline
Cut &  \multicolumn{2}{c}{No. Events}    \\ \hline
Triggered Events  &  \multicolumn{2}{c}{128,761}\\ 
Electron ID \& $N_{iso}=0$&  \multicolumn{2}{c}{58534}  \\     
$Z$ Rejection &  \multicolumn{2}{c}{50943} \\ 
$N_{jets}<3$ &  \multicolumn{2}{c}{50415}  \\ 
Fiducial Jet &  \multicolumn{2}{c}{9097}  \\ 
Jet $E_T>10$ GeV/$c^2$ & \multicolumn{2}{c}{6478}    \\ 
$\ge 1$ Tau Seed &  \multicolumn{2}{c}{1265}   \\
$\Delta \phi>1.5$ & \multicolumn{2}{c}{1117}  \\ \hline 
 & \hspace{0.1in}$\sin{\Delta \phi}<0.3$\hspace{0.1in} & \hspace{0.1in}$\sin{\Delta \phi}>0.3$\hspace{0.1in}  \\ \hline 
 $\sin{\Delta \phi}$ & 510 & 607 \\ 
\# Tracks & 189 & 146  \\ 
Elect. Reject  &  98 & 93 \\ 
$m_{\tau}<2$ GeV/$c^2$& 93 & 84 \\ 
\# CES & 80 & 72 \\ 
Jet Width & 64 & 54  \\ 
$|Q|=1$ & 48 & 39   \\ 
Opp. Sign & 39 & 28   \\ 
$E_{l}\hspace{-0.13in}/\hspace{0.1in}>0$, $E_{h}\hspace{-0.16in}/\hspace{0.1in}>0$ & NA  & 8 \\ \hline \hline
\end{tabular}
\caption{\label{tab:summaryofcuts1}  \sl A summary of the cuts imposed and 
the number of events remaining in the data sample after each 
successive cut. }
\end{center}
\end{table}

\begin{table}
\begin{center}
\begin{tabular}{lcccc} \hline \hline
Cut & \multicolumn{4}{c}{Efficiency (\%)}     \\ \hline
 & \multicolumn{2}{c}{$Z \rightarrow \tau \tau$}&\multicolumn{2}{c}{$A^0+h^0 \rightarrow \tau \tau$ }\\ 
BR($\tau \rightarrow e$) & \multicolumn{2}{c}{32.5}  & \multicolumn{2}{c}{32.5}  \\ 
$|\eta_e|<1.2$ & \multicolumn{2}{c}{19.8} &  \multicolumn{2}{c}{26.5}    \\  
Electron ID \& $N_{iso}=0$& \multicolumn{2}{c}{35.6} &  \multicolumn{2}{c}{37.5}    \\     
L2 Trigger & \multicolumn{2}{c}{90.8}  & \multicolumn{2}{c}{91.0} \\ 
$Z$ Rejection & \multicolumn{2}{c}{93.3}  & \multicolumn{2}{c}{90.3}   \\ 
$N_{jets}<3$ & \multicolumn{2}{c}{97.7}  & \multicolumn{2}{c}{96.4}   \\ 
Fiducial Jet & \multicolumn{2}{c}{35.3}  & \multicolumn{2}{c}{40.6}   \\ 
Jet $E_T>10$ GeV & \multicolumn{2}{c}{89.4}  & \multicolumn{2}{c}{90.3}    \\ 
$\ge 1$ Tau Seed & \multicolumn{2}{c}{64.0}  & \multicolumn{2}{c}{65.1}    \\
$\Delta \phi>1.5$ & \multicolumn{2}{c}{100.0}  & \multicolumn{2}{c}{100.0}  \\ \hline 
 & \multicolumn{2}{c}{$\sin{\Delta \phi}<0.3$} & \multicolumn{2}{c}{$\sin{\Delta \phi}>0.3$}  \\ \hline
% & BTB \hspace{0.05in}&\hspace{0.05in} nBTB\hspace{0.1in} &\hspace{0.1in} BTB \hspace{0.05in}& \hspace{0.05in}nBTB  \\ \hline
 $\sin{\Delta \phi}$ & 84.1 &  15.9 & 80.7  & 19.4 \\ 
\# Tracks & 84.3 & 85.9 & 82.3   & 83.0  \\ 
Elect. Reject  & 95.5  & 96.0 & 95.2   & 95.0 \\ 
%Elect. Reject (Ele) & 1.1 & 2.0 & 1.4  & 0.8 \\ 
$m_{\tau}<2$ GeV & 98.5 & 98.6 & 98.5  & 98.4 \\ 
\# CES & 97.0  & 96.6 & 96.9 & 97.7 \\ 
Jet Width & 92.2 & 92.5& 92.7  & 92.0 \\ 
$|Q|=1$ & 93.1 & 92.9 & 92.7  & 92.8 \\ 
Opp. Sign & 99.8 & 100.0 & 99.8  & 100.0 \\ 
$E_{l}\hspace{-0.13in}/\hspace{0.1in}>0$, $E_{h}\hspace{-0.16in}/\hspace{0.1in}>0$ & NA & 56.8& NA  & 58.0 \\ \hline \hline
\end{tabular}
\caption{\label{tab:ztt_eff1} Efficiency of each cut imposed on $Z \rightarrow \tau \tau$ and $A^0+h^0 \rightarrow \tau \tau$
simulated events.  Back-to-back and non-back-to-back events 
are denoted $\sin{\Delta \phi}<0.3$ and $\sin{\Delta \phi}>0.3$, 
respectively.   }
\end{center}
\end{table}

%Table~\ref{tab:summaryofcuts1}  
Table~III
summarizes the cuts that we impose
and the number of events remaining in the sample after each cut.
%Table~\ref{tab:ztt_eff1} 
Table~IV
summarizes the efficiency of each cut on
the $Z\rightarrow \tau \tau$ and $A^0+h^0 \rightarrow \tau \tau$
simulated events.

\section{Backgrounds \label{chap:ANA2}}

\label{sec:fakes}

The backgrounds can be 
classified into two categories: physics and 
instrumental backgrounds.  
The former includes 
$Z \rightarrow \tau \tau$ and $Z/\gamma^{*} \rightarrow ee$.  
%Instrumental backgrounds come from events in which leptons 
%are faked by QCD jets.  Either the electron candidate or the 
%hadronic tau candidate, or both, may be faked by a jet, for
%example in $W$+jets events or QCD events.  
%In some QCD events, a real electron from a 
%conversion pair may be identified as the
%$\tau_{e}$ candidate, while a recoil jet fakes a hadronic tau.  
%From $W(\rightarrow e\nu)$+jets events, we may also observe a real
%electron and a fake hadronic tau from a recoil jet.
%Some QCD events may contain both a fake tau and a fake electron.
%MC studies on $W(\rightarrow \tau \nu)$+jets have shown that it is
%an insignificant background. 
The latter fall into three categories:
events containing a 
conversion pair and a recoil jet, 
$W(\rightarrow e\nu)$+jets events and events containing
a jet that fakes an electron.  All of these contain a fake hadronic 
tau.  
We model signal and physics backgrounds 
using
{\sc Pythia} Monte-Carlo and the CDF detector simulation~\cite{Pythia}.

%\subsection{Fakes}

We estimate the
rate of all fake tau contributions combined instead
of estimating each source of fakes separately.
This way, we do not rely on modeling of jets, nor
on limited control samples for each separate
background.
We estimate the number of fakes expected to pass all of the cuts
using the fake rate technique described next.

%from the same data sample in which we perform the
%analysis.  We do this by folding in tau fake rates (measured from
%separate control samples) with jet multiplicities 
%observed in the data sample at a stage
%in the analysis where the data is 
%expected to be dominated by fake tau backgrounds.

We use five different control samples from CDF data to measure the
rate at which a jet fakes a hadronic tau by passing the tau identification
cuts for this analysis.
They are: (i) a sample dominated by events where a photon produced an
electron pair, (ii) a sample dominated by $W (\rightarrow e \nu)$+jets,
(iii) another dominated by $W (\rightarrow \mu\nu)$+jets and
two samples of events collected using inclusive triggers with
a jet satisfying (iv) $E_T>20$ GeV and (v) $E_T>50$ GeV.
The first three are classified as lepton samples and the latter
two are jet samples.
%the Conversion Sample,
%the $W (\rightarrow e \nu)$+jets Sample, the $W (\rightarrow \mu\nu)$+jets
%Sample and the Jet20 (Jet50) Sample, the latter a sample of events 
%triggered by a jet with $E_T>20$ GeV ($E_T>50$ GeV).
%These control samples are classified as lepton samples (the Conversion Sample,
%the $W (\rightarrow e \nu)$+jets Sample and the $W (\rightarrow \mu\nu)$+jets Sample)
%and jet samples (Jet20 Sample and the Jet50 Sample).

\begin{figure}
\scalebox{0.5}{\includegraphics[width=\textwidth]{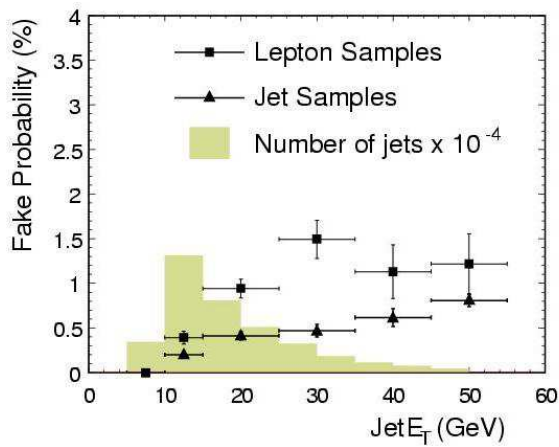}}
\caption{\label{fig:fakerates_lep_jet} Hadronic tau fake rates as measured in the lepton control
samples compared with that measured in the jet control samples.}
\end{figure}

A fake rate is defined as the probability that a jet passes
the hadronic tau identification requirements as described in
Sec.~\ref{sec:tauid}.
For all data control samples,
we measure the fake rates in bins defined by the 
$E_T$ of the jet associated with
the tau candidate.
Figure~\ref{fig:fakerates_lep_jet} shows the fake 
rates measured from
all three lepton samples combined and from both jet samples combined.
We measure fake rates less than about 1\%
from the jet samples, which are as good as in previous tau
analyses~\cite{leslie}.  
We find that the fake rates measured from the lepton
samples are approximately a factor of 2 higher than from the jet
samples.  
Two of the lepton samples are dominated by W+jets events where 
the recoil jet comes from a quark; quark jets are narrower,
have lower multiplicity and are thus more likely to pass the
tau identification cuts~\cite{quarkjetsgluonjets}.
Still, the reason for
this difference is not definitively understood.  
Since
the analysis is performed in a lepton sample, we use the fake rates
measured from the lepton samples as the central value and
the difference between the rates from the two different types of 
samples as a systematic uncertainty.
The histogram in Fig.~\ref{fig:fakerates_lep_jet} shows the
$E_T$ distribution of jets in the data sample just before tau
identification cuts are applied.  The fake rates are folded into this
$E_T$ distribution to predict the rate of fake tau background.

We find that fake taus from the $W$+jets samples are opposite-sign
from the lepton in the event (67.1 $\pm$ 3.0)\% of the time.  
This is due to a correlation between the W and the recoiling quark.  
The isolation cut enhances this correlation due to charge conservation.
In the conversion sample, the opposite-sign requirement is 
consistent with being 50\% efficient.
We take the central value of the 
efficiency for the opposite-sign cut to be the average 
of the two 
($(67.1 + 50)/2 = 58.6\%$) 
%$(67 + 50)/2 = 59\%$) 
and 
equally likely to be anywhere between 
50.0\% and 67.1\%.

To estimate the number of fake taus expected to pass all of
the cuts, we use the data sample with the following cuts removed:
(i) tau ID cuts (ii) the opposite sign requirement and
(iii) $E\hspace{-0.1in}/_{\tau_1,\tau_2}>0$. 
Then,
we fold in the measured fake rates with the $E_T$ spectrum
of jets.
There are 6478 events that pass these cuts and which contain at least 
one jet passing the fiducial cuts.   We apply our measured fake rates to 
6972 jets from these events.  For each jet, we give it a weight equal
to the fake rate measured for its $E_T$.  
%For each event, the probability
%that an event contains a fake is the opposite of the probability that
%no jet fakes a tau.  
We expect 21.0 $\pm$ 12.1 back-to-back and 
29.8 $\pm$ 16.8 non-back-to-back for a total of 
50.8 $\pm$ 29.0 fakes before the 
remaining cuts are applied.  

We apply two final cuts to improve the purity of the sample.
We apply the opposite sign requirement, 
taking the efficiency to be 58.6\%.  Including the systematic
error from this cut, we expect 10.5 $\pm$ 6.0 back-to-back and
14.9 $\pm$ 8.5 non-back-to-back events at this stage.
The final cut is the $E\hspace{-0.1in}/_{\tau_1,\tau_2}>0$ cut applied to the 
non-back-to-back events only.  This is measured from the same
background-dominated sample, subtracting out $Z \rightarrow \tau \tau$
and $Z \rightarrow ee$ contamination.
We find that
this
cut removes 89.0\% of the fake tau background.  This brings the number of 
predicted non-back-to-back events with a fake tau to 1.6.  Added to the back-to-back
events, we expect 1.6 + 10.5 = 12.1 events containing a fake hadronic tau 
to pass the analysis cuts.  At each stage of this estimation, we
account for the 10\%-level 
$Z \rightarrow \tau \tau$ and $Z \rightarrow ee$ contamination
of the background-dominated sample.

\section{Summary of Systematic Uncertainties \label{section-systematics}}

\label{sec:systematics}

In 
%Tab.~\ref{tab:systematics2}
Tab.~V
, we summarize the systematics
on the backgrounds and signal.  $Z \rightarrow ee$ is not included in the table because
the expected rate is based on a low number of background events.  We
expect 0.6 $\pm$ 0.3 $Z \rightarrow ee$ events in the counting experiment.
The systematics
on the fake tau background are described in Sec.~\ref{sec:fakes}.

The error on the Run 1b luminosity at CDF is 4.1\%~\cite{luminosity}.
The systematic error on the electron identification cuts is
1.8\%, as noted in Sec.~\ref{section-cuts}.  This error comes
from the limited size of the $Z\rightarrow ee$ sample used for
comparing the simulation with data.
The systematic error due to the modeling
of the trigger efficiency is obtained by
moving the parameters in the energy and $\eta$-dependent 
efficiency by one standard deviation in each
direction 
%(such that they all move the efficiency in the same
%direction) 
and measuring the effect on the total
efficiency of all cuts.  
This systematic is 1.6\% (1.7\%) for $A^0/h^0 \rightarrow \tau \tau$ ($Z \rightarrow \tau \tau$).

The one systematic uncertainty on the yield of signal events that varies significantly with the
mass of the Higgs boson is the uncertainty on the cross section due to the low $\sqrt{Q^2}$ tail.
Here are the systematic uncertainties due to this effect for each parameter space point considered
in this note:  
$m_{A^0}=100$~GeV/c$^2$, $\tan{\beta}=50$, 0.5\%;  
$m_{A^0}=110$~GeV/c$^2$, $\tan{\beta}=50$, 2.5\%; $m_{A^0}=120$~GeV/c$^2$, 
$\tan{\beta}=50$, 3.5\%, 
$m_{A^0}=140$~GeV/c$^2$, $\tan{\beta}=50$, 7.2\%; $m_{A^0}=140$~GeV/c$^2$, $\tan{\beta}=80$, 
21.3\%.
%Since the CDF measured cross section is used to
%scale the $Z$ MC samples to the expected rate at CDF, the systematic
%uncertainty on that measurement gives us an uncertainty on the $Z$ cross sections used.
The systematic uncertainty on the $Z$ production cross section 
is $1.7\%$ based on the CDF measurement~\cite{zprod}.
%for each of 
%$Z/\gamma^* \rightarrow \tau \tau$ and
%$Z/\gamma^* \rightarrow ee$.

By studying isolated pions, we have shown that
the detector simulation 
overestimates the width of jets from charged pions, and thus
underestimates the efficiency of the cut on jet width
on the hadronic tau candidates.  Therefore, we take the 
efficiency for that cut to be the average of the simulated 
efficiency and 100\%, which is (100.0+92.2)/2 = 96.1\%.
The systematic error on this cut is half the difference
between the efficiency from simulation 
and 100\%, which is (100.0-92.2)/2 = 3.9\%.
The
electron rejection cut applied to the hadronic tau candidate
is not modeled sufficiently for hadronic decays.
The same study of isolated pions previously mentioned showed that 
the hadronic energy deposited
by charged pions is underestimated by the simulation, so that
the efficiency of this cut is underestimated.  
We take the efficiency to be the average of the 
efficiency from simulation for hadronic taus and 100\%, which is (100.0+95.5)/2 = 97.8\%.
The systematic error on this cut is half the difference
between the simulated efficiency and 100\%, which is (100.0-95.5)/2 = 2.2\%.
  We obtain
the jet energy scale systematic by varying the energies of all of the
jets (except those identified as electrons) up and down by 5\%.  
The resulting systematic
error is 1.0\% (1.2\%) 
for $A^0/h^0 \rightarrow \tau \tau$ ($Z \rightarrow \tau \tau$).
There is a systematic uncertainty attributed to the tau fake rates since
we measured different rates in lepton and jet samples.
We set this systematic uncertainty to the
the difference between the fake rates measured
 in the two types of samples;
it is the dominant systematic error at 57.1\%.

\begin{table} 
\begin{center}
\begin{tabular}{lcccc} \hline \hline
systematic uncertainty & $A^0/h^0 \rightarrow \tau \tau$ & $Z \rightarrow \tau \tau$ & Fakes\\
  & (\%) & (\%) & (\%) &    \\ \hline
luminosity & 4.1 & 4.1 & -- \\ 
cross section & 0.5 & 1.7 & -- \\ 
electron ID & 1.8 & 1.8 & -- \\ 
sample dependence of fake rates & -- & -- & 57.1 \\ 
opposite sign & -- & -- & 14.7 \\ 
jet width & 3.9 & 3.9 & -- \\ 
jet energy scale & 1.0 & 1.2 & -- \\ 
%PDF's & -- & -- & -- & -- \\ \hline
%ISR/FSR & -- & -- & -- & -- \\ \hline
trigger efficiency & 1.6 & 1.7 & -- \\ 
electron rejection & 2.2 & 2.2 & -- \\ \hline 
total error (\%) & 6.7 & 7.0 & NA \\ \hline \hline 
\end{tabular}
\caption{\label{tab:systematics2} \sl Summary of systematic uncertainties 
on the counting experiment for the final cuts.  Here we refer to
 Higgs boson events corresponding to the parameter space point
$m_{A^0}=100$ GeV, $\tan{\beta}=50$.
All systematic uncertainties
are quoted as a percentage of the total number of events observed after all cuts are applied.  $Z\rightarrow ee$ is not included in the table because it is based on a low number of MC events.  The errors attributed to the rate of fake taus are not both Gaussian, and are therefore not added in quadrature.  See the text. 
}
\end{center}
\end{table}

\section{Results \label{section-results}}

We set limits on direct $A^0/h^0$ production in
the MSSM based on a counting experiment using events from both the
back-to-back and non-back-to-back samples.  Then we show the
limits achieved from a binned likelihood fit to the
di-tau mass distribution from the non-back-to-back events alone.
Our nominal limits 
%on the MSSM parameters from the di-tau channel in this data set 
come from the counting experiment utilizing both back-to-back and
non-back-to-back events.

\subsection{All Events}

\begin{figure}
\scalebox{0.42}{\includegraphics{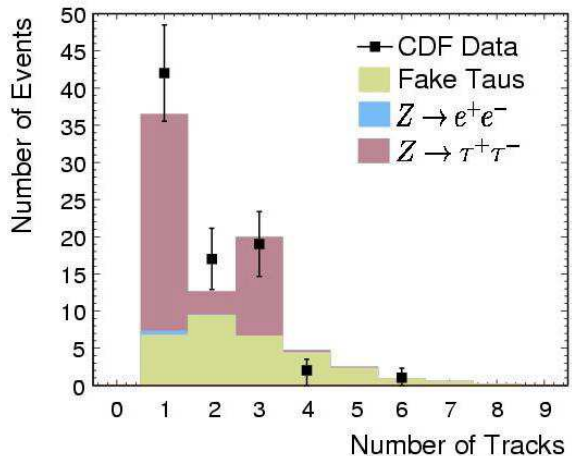}}
\caption{\label{fig:ntracks_counting} Track multiplicity distribution for the counting experiment,
but \emph{without} the $|Q|=1$, opposite sign and $N^{trks}_{cone}<4$ requirements. }
\end{figure}

We plot the track multiplicity of the tau candidates 
%in the
%events which pass our cuts because hadronic taus
%appear predominantly in the 1-track and 3-track bins.
%This plot allows us to compare the tau, fake composition
%in the data with the background estimates.  We would also 
%like to demonstrate that we do observe
%the irreducible background, $Z \rightarrow \tau \tau$.
%Before we look at the track multiplicity plot, 
%we 
after
imposing all of the analysis cuts \emph{except}
the following cuts:  i) $|\sum{Q_i}|=1$, ii) 
$N^{trks}_{cone}<4$ and iii) the opposite sign requirement.
Figure~\ref{fig:ntracks_counting} shows 
the number of tracks in the 0.175 cone around the tau seed
in the hadronic tau candidate in the event.
We expect 78.2 events to appear in this plot and we observe 81.
%Remember that the events that pass all of our analysis cuts 
%are a subset of the
%events that appear in Fig.~\ref{fig:ntracks_counting}.  Also
%remember that 
%the events in Fig.~\ref{fig:ntracks_counting} do not have the 
%three cuts listed above imposed, so the fake contribution is higher here
%than after all cuts are imposed.
Table~\ref{tab:ntracks_counting} lists the number of events expected
of each background type. 
%in Fig.~\ref{fig:ntracks_counting} and after
%the subsequent cuts are imposed.  
The data and the prediction show good
agreement at each stage.  Of the 47 final observed events, 35
are 1-track and 12 are 3-track.  The final three cuts reduce the
fake background in these two bins by nearly a factor of 2 compared to that
shown in Fig.~\ref{fig:ntracks_counting} and leaves the 
$Z\rightarrow\tau \tau$ background in those bins virtually unchanged.
\begin{table} 
\begin{center}
\begin{tabular}{cccccc} \hline \hline
& $Z \rightarrow \tau \tau$ & $Z \rightarrow ee$ & Fake Taus & Total & Obs. \\ \hline 
All cuts but  &   &  &  &  &  \\ 
$|Q|=1$, Opposite Sign &  46.1 & 0.56 & 31.5 & 78.2 & 81 \\ 
$|Q|=1$,\# Tracks$<4$ & 42.4 & 0.56  & 20.2 & 63.1 & 61       \\ 
Opposite Sign & 42.4 & 0.56  & 11.8 & 54.8 & 47 \\ \hline \hline
\end{tabular}
\caption{\label{tab:ntracks_counting}\sl Number of each background type expected, in the track multiplicity 
plot and after subsequent cuts are imposed, compared with the number observed. }
\end{center}
\end{table}

\subsection{Non-Back-to-Back Events}
\begin{figure}
\scalebox{0.42}{\includegraphics{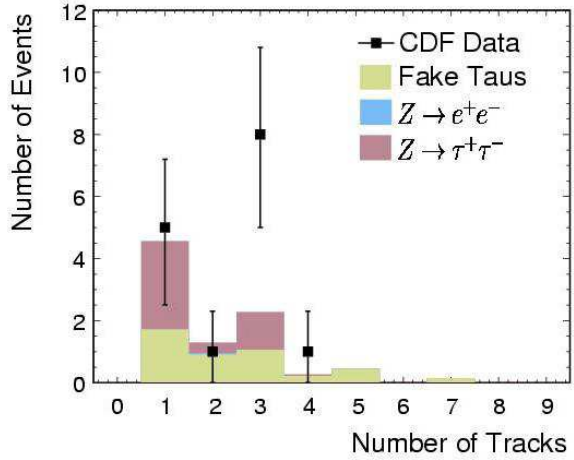}}
\caption{\label{fig:ntracks_nbtb}Track multiplicity distribution for the non-back-to-back events
\emph{with} the $E_{l,h}\hspace{-0.23in}/\hspace{0.12in}>0$ requirement,
but \emph{without} the $N^{trks}_{cone}<4$, 
$|Q|=1$ and opposite sign requirements. }
\end{figure}

Figure~\ref{fig:ntracks_nbtb} shows the track multiplicity distribution
for the non-back-to-back events only (since we will be
performing the di-tau mass reconstruction on these events).
These events have passed the $E_{l,h}\hspace{-0.23in}/\hspace{0.12in}>0$
requirement but the $|Q|=1$, opposite sign, and $N^{trks}_{cone}<4$ 
requirements have not been imposed.
We expect 9.2 events to appear in the plot and we observe 15.
The probability for the disagreement between data and the prediction
to be equal to or more than what was observed is 3.2\%.  
Table~\ref{tab:ntracks_nbtb} lists the number of events expected
of each background type in Fig.~\ref{fig:ntracks_nbtb} including
non-back-to-back events only.  We also show the expectation compared
to the observed after
the subsequent cuts are imposed.  The final 8 events contain 5
1-track events and 3 3-track events.
\begin{table} 
\begin{center}
\begin{tabular}{cccccc} \hline \hline 
& $Z \rightarrow \tau \tau$ & $Z \rightarrow ee$ & Fake Taus & Total & Obs. \\ \hline 
All cuts but &   &  &  &  &   \\ 
$|Q|=1$, Opposite Sign & 4.4  & 0.05 & 4.5 & 9.2 & 15  \\ 
$|Q|=1$,\# Tracks$<4$ & 4.1 & 0.05  & 3.0 & 7.1 & 13       \\ 
Opposite Sign & 4.1 & 0.05 & 1.7 & 5.9 & 8 \\ \hline \hline
\end{tabular}
\caption{\label{tab:ntracks_nbtb}\sl Number of each background type expected, in the track multiplicity 
plot and after subsequent cuts are imposed, compared with the number observed.  Only non-back-to-back events are included. }
\end{center}
\end{table}

\subsection{Limits}

The numbers of observed events do not show an excess above
the standard model expectation for background processes.
%indicate an excess with respect to SM process.
Therefore, we set a limit on the product of the 
cross section and branching ratio of $A^0/h^0 \rightarrow \tau \tau$
($\sigma \cdot$ BR) at 95\% confidence level.
We take the branching ratio to be 9\% for all parameter space points
considered.
We use a Bayesian method with a flat prior 
based on a likelihood
that is smeared to 
account for systematic errors.

%\begin{figure}
%\scalebox{0.42}{	\includegraphics{likelihood.eps}}
%\caption{\label{fig:likelihood}Likelihood distribution for the counting experiment
%at $m_{A^0}=100$~GeV/c$^2$, $\tan{\beta}=50$, considering statistical errors
%only (solid) and after systematics smear the distribution (dashed).
%The arrows point to where the integral under each curve reaches 0.95. }
%\end{figure}

Table~\ref{tab:limit_vs_mass} 
shows the predicted and observed 
upper limits on $\sigma \cdot$ BR
 for $A^0/h^0$ production in the MSSM as a function
of $m_{A^0}$ at $\tan{\beta}=50$.  
For $m_{A^0}=100$~GeV/c$^2$, $\tan{\beta}=50$ in the MSSM,
 the expected limit is 23.4 signal
events, corresponding to 102 pb.
Since we observed slightly fewer events than we expected, the
observed limits are better than our expected limits.  The observed limit
is 77.9 pb.  
The limits on the $\sigma \cdot$ BR improve
with increasing mass since the efficiency improves, but we are less
sensitive to the MSSM theory at higher mass due to the steeply falling
predicted cross section.
\begin{table} 
\begin{center}
\begin{tabular}{ccccccc} \hline \hline 
Mass  & $\sigma \cdot$BR & Eff. & 95\% CL& 95\% CL \\ 
GeV/c$^2$ & pb & \% &  Exp. (pb) & Obs. (pb)       \\ \hline
100 & 11.0  & 0.78 & 102 & 77.9 \\ 
110 & 5.8 & 0.91 & 87.4 & 67.2 \\ 
120 & 4.0 & 0.97 & 82.3 & 63.0 \\  
140 & 1.6 & 1.1 & 75.2 & 57.9 \\ \hline \hline                  
\end{tabular}
\caption{\label{tab:limit_vs_mass}
95\% CL limits vs. $m_{A^0}$ for $\tan{\beta}=50$. The first two 
columns are the mass of the Higgs boson for each parameter point and
the corresponding theoretically predicted $\sigma \cdot$ BR.}
\end{center}
\end{table}

\begin{figure}
\scalebox{0.4}{\includegraphics{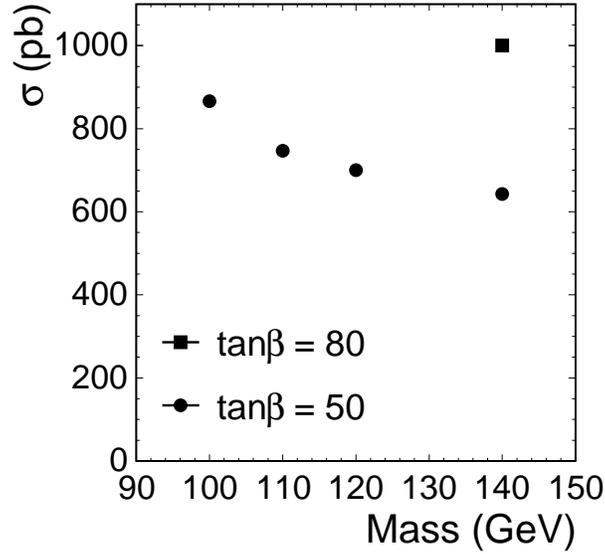}}
\caption{\label{fig:limits}Summary of limits achieved in this analysis 
on the cross section for directly produced Higgs bosons for each parameter space
point considered.}
\end{figure}

The cross section for producing $A^0/h^0$ in the MSSM scales
with $(\tan{\beta})^2$.
% it is tempting to scale the sensitivity with
%1/$(\tan{\beta})^2$.  However, this would be a mistake.  
The sensitivity does not improve by that same factor,
however, because the Higgs boson width scales as
$(\tan{\beta})^2$,
and the tail at
low $\sqrt{s}$ also becomes more prominent with increasing 
$\tan{\beta}$, increasing the
systematic error due to the uncertainty in the cross section in this region. 
At $m_{A^0}=140$~GeV/c$^2$ and $\tan{\beta}=80$,
the uncertainty on the efficiency of the selection of 
Higgs boson events due to this low mass tail
is 20\%, compared to 7.2\% at $m_{A^0}=140$~GeV/c$^2$, $\tan{\beta}=50$.
Also, both effects bring down the efficiency at higher $\tan{\beta}$:  
at $m_{A^0}=140$~GeV/c$^2$ and  $\tan{\beta}=80$, the efficiency is similar to 
the efficiency at a lower mass point:  $m_{A^0}=110$~GeV/c$^2$, $\tan{\beta}=50$.
Table~\ref{tab:limit_vs_tanb}
shows the limits for two different values of $\tan{\beta}$ for
$m_{A^0}=140$~GeV/c$^2$.  These limits are also summarized in Figure~\ref{fig:limits}.
\begin{table} 
\begin{center}
\begin{tabular}{ccccccc} \hline \hline 
$\tan{\beta}$  & $\sigma \cdot$ BR & Eff.  & 95\% CL &95\%CL \\ 
           & pb & \% &  Exp. (pb) & Obs. (pb)        \\ \hline
50 & 17.8 & 1.1  & 75.2 & 57.9\\ 
80 & 65.9 & 0.85 & 118 & 90 \\ \hline \hline 
\end{tabular}
\caption{\label{tab:limit_vs_tanb}\sl 95\% CL limits vs. $\tan{\beta}$ for $m_{A^0}=140$~GeV/c$^2$.  All limits are quoted in pb.  The efficiencies shown do not include branching
ratios.  The first two 
columns are the mass of the Higgs boson for each parameter point and
the corresponding theoretically predicted $\sigma \cdot$ BR.}
\end{center}
\end{table}

\begin{figure}
\scalebox{0.4}{\includegraphics{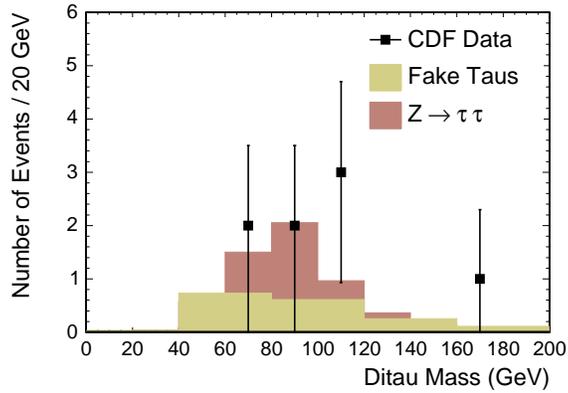}}
\caption{ \label{fig:massplot} Di-tau mass reconstruction for events in the non-back-to-back
region.  It is compared with the background prediction, shown for
fake tau backgrounds and then the $Z \rightarrow \tau \tau$ background
added to it.  The negligible contribution from $Z \rightarrow ee$ 
is not shown.}
\end{figure}

In the non-back-to-back region, after all cuts are applied,
we expect 5.9 events and observe 8.  Figure~\ref{fig:massplot}
shows the di-tau mass distribution for these 8 events compared
with the expectation.  In this region, limits are obtained by fitting
the mass distribution using a binned likelihood, 
with 14 bins between 60 GeV/c$^2$ and 200 GeV/c$^2$ in di-tau mass.

%Table~\ref{tab:limit_vs_mass_nbtb} 
Table~X 
shows the upper limits on the
MSSM $\sigma \cdot$ BR obtained from
the binned likelihood for four different values of $m_{A^0}$ for 
$\tan{\beta}=50$.  
%Again, the limits on the cross section do improve
%with increasing mass since the efficiency improves, but we are less
%sensitive to the MSSM theory at higher mass due to the steeply falling
%predicted cross section.  
At $m_{A^0}=100$~GeV/c$^2$, 
where
the Higgs boson mass is nearly on top of the $Z$ mass, the mass
reconstruction is less effective than at higher masses, so 
the expected limit from the binned likelihood in the non-back-to-back region 
is approximately 2.4 times worse than the 
expected limit
from the counting experiment without the non-back-to-back requirement.  
At $m_{A^0}=140$~GeV/c$^2$, which is
approximately 2 RMS away from the $Z$ in the di-tau mass variable,
the expected limit from the binned likelihood using the 
non-back-to-back events 
is 2.1 times worse than the 
counting experiment limit, showing a modest improvement, but still
not coming close to the expected limit from the counting experiment.
With more data collected in Run 2, the power of the di-tau mass reconstruction 
technique will improve.
%Table~\ref{tab:limit_vs_tanb_nbtb} 
Table~XI
shows the upper limits obtained from
the binned likelihood for two different values of $\tan{\beta}$.  
%Again, we cannot scale the sensitivity with $(\tan{\beta})^2$ because the
%efficiency is degraded at higher $\tan{\beta}$.

\begin{table} 
\begin{center}
\begin{tabular}{ccccc} \hline \hline 
Mass  & $\sigma \cdot$ BR & Eff. & 95\% CL& 95\% CL \\ 
GeV/c$^2$ & pb & \% &  Exp. (pb) & Obs. (pb)       \\ \hline
100 & 11.0  & 0.093 & 247 & 395  \\ 
110 & 5.8 & 0.10 & 218 & 379 \\ 
120 & 4.0 & 0.105 & 199 & 363 \\  
140 & 1.6 & 0.12 & 158 & 309 \\ \hline \hline                  
\end{tabular}
\caption{\label{tab:a}
95\% CL limits vs. $m_{A^0}$ for $\tan{\beta}=50$, using the binned likelihood. The first two 
columns are the mass of the Higgs boson for each parameter point and
the corresponding theoretically predicted $\sigma \cdot$ BR.}
\end{center}
\end{table}

\begin{table} 
\begin{center}
\begin{tabular}{ccccc} \hline \hline 
$\tan{\beta}$  & $\sigma \cdot$ BR & Eff.&95\% CL & 95\% CL \\ 
 & pb  & (\%) &  Exp. (pb) & Obs. (pb)    \\ \hline
50 & 1.6 &  0.12 & 158 & 309 \\ 
80 & 5.9 & 0.09 & 274 & 533 \\ \hline \hline
\end{tabular}
\caption{\label{tab:limit_vs_tanb_nbtb} 95\% CL limits vs. $\tan{\beta}$ for $m_{A^0}=140$~GeV/c$^2$, using the binned likelihood.  All limits are quoted in pb.  The efficiencies shown do not include branching
ratios.  The first two 
columns are the mass of the Higgs boson for each parameter point and
the corresponding theoretically predicted $\sigma \cdot$ BR.}
\end{center}
\end{table}

\label{chap:CONCLUSION}

\section{Conclusions}

We have performed a search for directly produced 
Higgs bosons decaying to two taus where one tau decays
to an electron and the other hadronically 
in Run 1b data at CDF.  This is
the first Higgs boson search based on the di-tau final
state at a hadron collider.
The number of events that pass all of the cuts
 is consistent with the background expectation.
This agreement between
data and background demonstrates our capability to reconstruct 
$Z\rightarrow \tau \tau$ final states, the irreducible background
for this analysis.
At a benchmark parameter space point, $m_{A^0}=100$~GeV/c$^2$ and 
$\tan{\beta}=50$,
we are sensitive to a $\sigma \cdot$ BR of 102 pb
compared to the 11.0 pb predicted in the MSSM.
%9.2 times that predicted in 
%the MSSM (1130 pb ).  
The observed
limit at this parameter space point is 77.9~pb.  
%The limit on the 
%cross section improves with increasing mass.
 %because the efficiency improves at higher mass.  
%However, the sensitivity to the theory 
%at higher mass is worse because the
%cross section falls  %so 
%steeply.

A di-tau mass reconstruction is performed for tau candidate
pairs which are not back-to-back, for the first time with hadron
collider data.
%Using events where the tau candidates are not back-to-back, 
%a di-tau mass reconstruction is performed for the first time in
%hadron collider data.  
%However, t
The modest sensitivity that one gains
from a limit binned in mass is not nearly enough to make up for
the hit in efficiency taken when only 
non-back-to-back events are considered.  At $m_{A^0}=100$~GeV/c$^2$, the binned mass
limit at $\tan{\beta}=50$ from non-back-to-back events alone 
is 
395 pb. %2.4 times worse than from the full counting experiment.
At $m_{A^0}=140$~GeV/c$^2$, $\tan{\beta}=50$, the binned mass limit is  
309 pb, %2.1 times worse
 showing a modest improvement as the Higgs boson mass 
is moved away from the $Z$ mass.

While this search does not
have the sensitivity to the Standard Model Higgs boson
that prior searches using decays to pairs of $b$ quarks,
it lays the groundwork
for
%While the search does not match the sensitivity of preceding Higgs boson searches, it does lay the groundwork for 
similar analysis to be performed by future experiments.  The di-tau mass reconstruction
technique demonstrated here may also be useful for searches for other processes where a Higgs boson
is produced with a recoil, such as $Hb$ or $Hb\bar{b}$. 

\section{Acknowledgements \label{section-acknowledgements}}

We thank the Fermilab staff and the technical staffs of the participating institutions for their
vital contributions.  This work was supported by the U.S. Department of Energy and National
Science Foundation; the Italian Istituto Nazionale di Fisica Nucleare; the Ministry of Education,
Science and Culture of Japan; the Natural Sciences and Engineering Research Council of Canada;
the National Science Council of the Republic of China; the A.P. Sloan Foundation.

\newpage
\bibliography{Ahtautau_prd_draft1}% Produces the bibliography via BibTeX.

\begin{thebibliography}{39}
\expandafter\ifx\csname natexlab\endcsname\relax\def\natexlab#1{#1}\fi
\expandafter\ifx\csname bibnamefont\endcsname\relax
  \def\bibnamefont#1{#1}\fi
\expandafter\ifx\csname bibfnamefont\endcsname\relax
  \def\bibfnamefont#1{#1}\fi
\expandafter\ifx\csname citenamefont\endcsname\relax
  \def\citenamefont#1{#1}\fi
\expandafter\ifx\csname url\endcsname\relax
  \def\url#1{\texttt{#1}}\fi
\expandafter\ifx\csname urlprefix\endcsname\relax\def\urlprefix{URL }\fi
\providecommand{\bibinfo}[2]{#2}
\providecommand{\eprint}[2][]{\url{#2}}

\bibitem[{\citenamefont{Nilles}(1984)}]{mssm1}
\bibinfo{author}{\bibfnamefont{H.}~\bibnamefont{Nilles}},
  \bibinfo{journal}{Phys. Rep.} \textbf{\bibinfo{volume}{110}},
  \bibinfo{pages}{1} (\bibinfo{year}{1984}).

\bibitem[{\citenamefont{Haber and Kane}(1985)}]{mssm2}
\bibinfo{author}{\bibfnamefont{H.}~\bibnamefont{Haber}} \bibnamefont{and}
  \bibinfo{author}{\bibfnamefont{G.}~\bibnamefont{Kane}},
  \bibinfo{journal}{Phys. Rep.} \textbf{\bibinfo{volume}{117}},
  \bibinfo{pages}{75} (\bibinfo{year}{1985}).

\bibitem[{\citenamefont{Barbieri}(1988)}]{mssm3}
\bibinfo{author}{\bibfnamefont{R.}~\bibnamefont{Barbieri}},
  \bibinfo{journal}{Riv. Nuovo Cim.} \textbf{\bibinfo{volume}{11
  $n^{\circ}$4}}, \bibinfo{pages}{1} (\bibinfo{year}{1988}).

\bibitem[{\citenamefont{J.~Ellis}(1991)}]{mssmhiggs1}
\bibinfo{author}{\bibfnamefont{F.~Z.} \bibnamefont{J.~Ellis},
  \bibfnamefont{G.~Ridolfi}}, \bibinfo{journal}{Phys. Lett.}
  \textbf{\bibinfo{volume}{B257}}, \bibinfo{pages}{83} (\bibinfo{year}{1991}).

\bibitem[{\citenamefont{Y.~Okada}(1991)}]{mssmhiggs2}
\bibinfo{author}{\bibfnamefont{T.~Y.} \bibnamefont{Y.~Okada},
  \bibfnamefont{M.~Yamaguchi}}, \bibinfo{journal}{Prog. Theor. Phys.}
  \textbf{\bibinfo{volume}{85}}, \bibinfo{pages}{1} (\bibinfo{year}{1991}).

\bibitem[{\citenamefont{Haber and Hempfling}(1991)}]{mssmhiggs3}
\bibinfo{author}{\bibfnamefont{H.}~\bibnamefont{Haber}} \bibnamefont{and}
  \bibinfo{author}{\bibfnamefont{H.}~\bibnamefont{Hempfling}},
  \bibinfo{journal}{Phys. Rev. Lett.} \textbf{\bibinfo{volume}{66}},
  \bibinfo{pages}{1815} (\bibinfo{year}{1991}).

\bibitem[{lep()}]{lep_mssmhiggs}
\bibinfo{note}{LEP Higgs Working Group, Searches for the neutral Higgs bosons
  of the MSSM: Preliminary combined results using LEP data collected at
  energies up to 209~GeV, hep-ex/0107030 (2001).}

\bibitem[{\citenamefont{Affolder et~al.}(2001)}]{valls}
\bibinfo{author}{\bibfnamefont{T.}~\bibnamefont{Affolder}} \bibnamefont{et~al.}
  (\bibinfo{collaboration}{CDF Collaboration}), \bibinfo{journal}{Phys. Rev.
  Lett.} \textbf{\bibinfo{volume}{86}}, \bibinfo{pages}{4472}
  (\bibinfo{year}{2001}).

\bibitem[{\citenamefont{Acosta et~al.}(2004)}]{stops}
\bibinfo{author}{\bibfnamefont{D.}~\bibnamefont{Acosta}} \bibnamefont{et~al.}
  (\bibinfo{collaboration}{CDF Collaboration}), \bibinfo{journal}{Phys. Rev.
  Lett.} \textbf{\bibinfo{volume}{92}}, \bibinfo{pages}{051803}
  (\bibinfo{year}{2004}).

\bibitem[{\citenamefont{Abe et~al.}(1997{\natexlab{a}})}]{franklin}
\bibinfo{author}{\bibfnamefont{F.}~\bibnamefont{Abe}} \bibnamefont{et~al.}
  (\bibinfo{collaboration}{CDF Collaboration}), \bibinfo{journal}{Phys. Rev.
  Lett.} \textbf{\bibinfo{volume}{78}}, \bibinfo{pages}{2906}
  (\bibinfo{year}{1997}{\natexlab{a}}).

\bibitem[{\citenamefont{Abe et~al.}(1988)}]{cdf1}
\bibinfo{author}{\bibfnamefont{F.}~\bibnamefont{Abe}} \bibnamefont{et~al.}
  (\bibinfo{collaboration}{CDF Collaboration}), \bibinfo{journal}{Nucl.
  Instrum. Methods Phys. Res. A} \textbf{\bibinfo{volume}{271}},
  \bibinfo{pages}{387} (\bibinfo{year}{1988}).

\bibitem[{\citenamefont{Abe et~al.}(1994)}]{cdf2}
\bibinfo{author}{\bibfnamefont{F.}~\bibnamefont{Abe}} \bibnamefont{et~al.}
  (\bibinfo{collaboration}{CDF Collaboration}), \bibinfo{journal}{Phys. Rev. D}
  \textbf{\bibinfo{volume}{50}}, \bibinfo{pages}{2966} (\bibinfo{year}{1994}).

\bibitem[{\citenamefont{Amidei et~al.}(1994)}]{cdf3}
\bibinfo{author}{\bibfnamefont{D.}~\bibnamefont{Amidei}} \bibnamefont{et~al.},
  \bibinfo{journal}{Nucl. Instrum. Methods Phys. Res. A}
  \textbf{\bibinfo{volume}{350}}, \bibinfo{pages}{73} (\bibinfo{year}{1994}).

\bibitem[{\citenamefont{Azzi et~al.}(1995{\natexlab{a}})}]{cdf4}
\bibinfo{author}{\bibfnamefont{P.}~\bibnamefont{Azzi}} \bibnamefont{et~al.},
  \bibinfo{journal}{Nucl. Instrum. Methods Phys. Res. A}
  \textbf{\bibinfo{volume}{360}}, \bibinfo{pages}{137}
  (\bibinfo{year}{1995}{\natexlab{a}}).

\bibitem[{\citenamefont{Balka et~al.}(1988)}]{ces1}
\bibinfo{author}{\bibfnamefont{L.}~\bibnamefont{Balka}} \bibnamefont{et~al.},
  \bibinfo{journal}{Nucl. Instrum. Methods Phys. Res. A}
  \textbf{\bibinfo{volume}{268}}, \bibinfo{pages}{50} (\bibinfo{year}{1988}).

\bibitem[{\citenamefont{Hahn et~al.}(1988)}]{ces2}
\bibinfo{author}{\bibfnamefont{S.~R.} \bibnamefont{Hahn}} \bibnamefont{et~al.},
  \bibinfo{journal}{Nucl. Instrum. Methods Phys. Res. A}
  \textbf{\bibinfo{volume}{267}}, \bibinfo{pages}{351} (\bibinfo{year}{1988}).

\bibitem[{\citenamefont{Yasuoka et~al.}(1988)}]{ces3}
\bibinfo{author}{\bibfnamefont{K.}~\bibnamefont{Yasuoka}} \bibnamefont{et~al.},
  \bibinfo{journal}{Nucl. Instrum. Methods Phys. Res. A}
  \textbf{\bibinfo{volume}{267}}, \bibinfo{pages}{315} (\bibinfo{year}{1988}).

\bibitem[{\citenamefont{Wagner et~al.}(1988)}]{ces4}
\bibinfo{author}{\bibfnamefont{R.~G.} \bibnamefont{Wagner}}
  \bibnamefont{et~al.}, \bibinfo{journal}{Nucl. Instrum. Methods Phys. Res. A}
  \textbf{\bibinfo{volume}{267}}, \bibinfo{pages}{330} (\bibinfo{year}{1988}).

\bibitem[{\citenamefont{Devlin et~al.}(1988)}]{ces5}
\bibinfo{author}{\bibfnamefont{T.}~\bibnamefont{Devlin}} \bibnamefont{et~al.},
  \bibinfo{journal}{Nucl. Instrum. Methods Phys. Res. A}
  \textbf{\bibinfo{volume}{267}}, \bibinfo{pages}{24} (\bibinfo{year}{1988}).

\bibitem[{\citenamefont{Azzi et~al.}(1995{\natexlab{b}})}]{svxprime}
\bibinfo{author}{\bibfnamefont{P.}~\bibnamefont{Azzi}} \bibnamefont{et~al.},
  \bibinfo{journal}{Nucl. Instrum. Methods Phys. Res. A}
  \textbf{\bibinfo{volume}{360}}, \bibinfo{pages}{137}
  (\bibinfo{year}{1995}{\natexlab{b}}).

\bibitem[{\citenamefont{Sjostrand et~al.}(2001)}]{Pythia}
\bibinfo{author}{\bibfnamefont{T.}~\bibnamefont{Sjostrand}}
  \bibnamefont{et~al.}, \bibinfo{journal}{Computer Physics Commun.}
  \textbf{\bibinfo{volume}{135}}, \bibinfo{pages}{238} (\bibinfo{year}{2001}).

\bibitem[{hig()}]{higlu}
\bibinfo{note}{M. Spira, hep-ph/9510347 (1995).}

\bibitem[{isa()}]{isajet}
\bibinfo{note}{H. Baer, F. Paige, S. Protopopescu, and X. Tata, hep-ph/0001086
  (2000).}

\bibitem[{\citenamefont{Ellis et~al.}(1988)\citenamefont{Ellis, Hinchliffe,
  Soldate, and Bij}}]{ptcalc}
\bibinfo{author}{\bibfnamefont{R.}~\bibnamefont{Ellis}},
  \bibinfo{author}{\bibfnamefont{I.}~\bibnamefont{Hinchliffe}},
  \bibinfo{author}{\bibfnamefont{M.}~\bibnamefont{Soldate}}, \bibnamefont{and}
  \bibinfo{author}{\bibfnamefont{J.~V.~D.} \bibnamefont{Bij}},
  \bibinfo{journal}{Nucl. Phys. B} \textbf{\bibinfo{volume}{297}},
  \bibinfo{pages}{221} (\bibinfo{year}{1988}).

\bibitem[{\citenamefont{Eidelman et~al.}(2004)}]{PDG}
\bibinfo{author}{\bibfnamefont{S.}~\bibnamefont{Eidelman}}
  \bibnamefont{et~al.}, \bibinfo{journal}{Phys. Lett. B}
  \textbf{\bibinfo{volume}{592}}, \bibinfo{pages}{1} (\bibinfo{year}{2004}).

\bibitem[{\citenamefont{Affolder et~al.}(2000{\natexlab{a}})}]{zpt}
\bibinfo{author}{\bibfnamefont{A.}~\bibnamefont{Affolder}} \bibnamefont{et~al.}
  (\bibinfo{collaboration}{CDF Collaboration}), \bibinfo{journal}{Phys. Rev.
  Lett.} \textbf{\bibinfo{volume}{84}}, \bibinfo{pages}{845}
  (\bibinfo{year}{2000}{\natexlab{a}}).

\bibitem[{\citenamefont{Jadach et~al.}(1990)\citenamefont{Jadach, Kuhn, and
  Was}}]{tauola}
\bibinfo{author}{\bibfnamefont{S.}~\bibnamefont{Jadach}},
  \bibinfo{author}{\bibfnamefont{J.~H.} \bibnamefont{Kuhn}}, \bibnamefont{and}
  \bibinfo{author}{\bibfnamefont{Z.}~\bibnamefont{Was}},
  \bibinfo{journal}{Comput. Phys. Commun.} \textbf{\bibinfo{volume}{64}},
  \bibinfo{pages}{275} (\bibinfo{year}{1990}).

\bibitem[{\citenamefont{Amidei et~al.}(1988)}]{l1l2trigger}
\bibinfo{author}{\bibfnamefont{D.}~\bibnamefont{Amidei}} \bibnamefont{et~al.},
  \bibinfo{journal}{Nucl. Instrum. Methods Phys. Res. A}
  \textbf{\bibinfo{volume}{269}}, \bibinfo{pages}{68} (\bibinfo{year}{1988}).

\bibitem[{\citenamefont{Affolder et~al.}(2000{\natexlab{b}})}]{Zee_eff}
\bibinfo{author}{\bibfnamefont{T.}~\bibnamefont{Affolder}} \bibnamefont{et~al.}
  (\bibinfo{collaboration}{CDF Collaboration}), \bibinfo{journal}{Phys. Rev.
  Lett.} \textbf{\bibinfo{volume}{84}}, \bibinfo{pages}{845}
  (\bibinfo{year}{2000}{\natexlab{b}}).

\bibitem[{\citenamefont{Abe et~al.}(1995{\natexlab{a}})}]{johnwahl}
\bibinfo{author}{\bibfnamefont{F.}~\bibnamefont{Abe}} \bibnamefont{et~al.}
  (\bibinfo{collaboration}{CDF Collaboration}), \bibinfo{journal}{Phys. Rev.}
  \textbf{\bibinfo{volume}{D52}}, \bibinfo{pages}{2624}
  (\bibinfo{year}{1995}{\natexlab{a}}).

\bibitem[{\citenamefont{Abe et~al.}(1995{\natexlab{b}})}]{saltzberg}
\bibinfo{author}{\bibfnamefont{F.}~\bibnamefont{Abe}} \bibnamefont{et~al.}
  (\bibinfo{collaboration}{CDF Collaboration}), \bibinfo{journal}{Phys. Rev.}
  \textbf{\bibinfo{volume}{D52}}, \bibinfo{pages}{4784}
  (\bibinfo{year}{1995}{\natexlab{b}}).

\bibitem[{\citenamefont{Abe et~al.}(1997{\natexlab{b}})}]{previoustau}
\bibinfo{author}{\bibfnamefont{F.}~\bibnamefont{Abe}} \bibnamefont{et~al.}
  (\bibinfo{collaboration}{CDF Collaboration}), \bibinfo{journal}{Phys. Rev.
  Lett.} \textbf{\bibinfo{volume}{79}}, \bibinfo{pages}{357}
  (\bibinfo{year}{1997}{\natexlab{b}}).

\bibitem[{\citenamefont{Groer}(1998)}]{leslie}
\bibinfo{author}{\bibfnamefont{L.}~\bibnamefont{Groer}}, Ph.D. thesis,
  \bibinfo{school}{Rutgers University} (\bibinfo{year}{1998}).

\bibitem[{cms()}]{cms}
\bibinfo{note}{CMS Technical Proposal (1994), 191-192.}

\bibitem[{atl({\natexlab{a}})}]{atlas1}
\bibinfo{note}{ATLAS Detector and Physics Performance. Technical Design Report.
  Vol. 1, CERN-LHCC-99-14}.

\bibitem[{atl({\natexlab{b}})}]{atlas2}
\bibinfo{note}{ATLAS Detector and Physics Performance. Technical Design Report.
  Vol. 2, CERN-LHCC-99-15}.

\bibitem[{qua()}]{quarkjetsgluonjets}
\bibinfo{note}{D. Acosta et al. (CDF Collaboration), FERMILAB-PUB-04-113-E}.

\bibitem[{\citenamefont{Cronin-Hennessy
  et~al.}(2000)\citenamefont{Cronin-Hennessy, Beretvas, and
  Derwent}}]{luminosity}
\bibinfo{author}{\bibfnamefont{D.}~\bibnamefont{Cronin-Hennessy}},
  \bibinfo{author}{\bibfnamefont{A.}~\bibnamefont{Beretvas}}, \bibnamefont{and}
  \bibinfo{author}{\bibfnamefont{P.~F.} \bibnamefont{Derwent}}
  (\bibinfo{collaboration}{CDF Collaboration}), \bibinfo{journal}{Nucl.
  Instrum. Meth.} \textbf{\bibinfo{volume}{A443}}, \bibinfo{pages}{37}
  (\bibinfo{year}{2000}).

\bibitem[{\citenamefont{Affolder et~al.}(2000{\natexlab{c}})}]{zprod}
\bibinfo{author}{\bibfnamefont{T.}~\bibnamefont{Affolder}} \bibnamefont{et~al.}
  (\bibinfo{collaboration}{CDF Collaboration}), \bibinfo{journal}{Phys. Rev.
  Lett.} \textbf{\bibinfo{volume}{84}}, \bibinfo{pages}{845}
  (\bibinfo{year}{2000}{\natexlab{c}}).

\end{thebibliography}

\end{document}